%% file: main.tex
\documentclass[sigplan,nonacm]{acmart}

\setcopyright{none}

\usepackage[utf8]{inputenc}

\usepackage{graphicx}   
\usepackage{subcaption} 
\usepackage{enumitem}
\usepackage{tikz}

\definecolor{GreenLLMRow}{HTML}{E8F5E9} 

\newcommand{\circlednum}[1]{%
  \tikz[baseline=(char.base)]{
    \node[shape=circle,draw,fill=black,inner sep=1pt] (char) {\textcolor{white}{#1}};}}

\usepackage{booktabs}
\usepackage{siunitx}

\usepackage[most]{tcolorbox}

\tcbset{
  takeaway/.style={
    colback=gray!10,
    colframe=black!60,
    fonttitle=\bfseries,
    coltitle=black,
    boxrule=0.6pt,
    arc=2mm,
    left=3pt,
    right=3pt,
    top=3pt,
    bottom=3pt,
  }
}

\AtBeginDocument{%
  }

\begin{document}

\sisetup{detect-weight=true,detect-family=true} 
\title{GreenLLM: SLO-Aware Dynamic Frequency Scaling for Energy-Efficient LLM Serving}


\author{Qunyou Liu}
\email{qunyou.liu@epfl.ch}
\affiliation{%
  \institution{Embedded Systems Laboratory (ESL), École Polytechnique Fédérale de Lausanne (EPFL)}
  \streetaddress{Rte Cantonale}
  \city{Lausanne}
  \state{Vaud}
  \country{Switzerland}
  \postcode{1015}
}

\author{Darong Huang}
\email{darong.huang@epfl.ch}
\affiliation{%
  \institution{Embedded Systems Laboratory (ESL), École Polytechnique Fédérale de Lausanne (EPFL)}
  \streetaddress{Rte Cantonale}
  \city{Lausanne}
  \state{Vaud}
  \country{Switzerland}
  \postcode{1015}
}

\author{Marina Zapater}
\email{marina.zapater@heig-vd.ch}
\affiliation{%
  \institution{Institute of Reconfigurable \& Embedded Digital Systems (REDS), School of Engineering and Management Vaud, HES-SO University of Applied Sciences and Arts Western Switzerland}
  \city{Yverdon-les-Bains}
  \country{Switzerland}
  \postcode{1401}
}

\author{David Atienza}
\email{david.atienza@epfl.ch}
\affiliation{%
  \institution{Embedded Systems Laboratory (ESL), École Polytechnique Fédérale de Lausanne (EPFL)}
  \streetaddress{Rte Cantonale}
  \city{Lausanne}
  \state{Vaud}
  \country{Switzerland}
  \postcode{1015}
}


\begin{abstract}

Large Language Models (LLMs) are rapidly becoming the backbone of modern cloud services, yet their inference costs are dominated by energy consumption on GPUs. Unlike traditional GPU workloads, LLM inference consists of two distinct stages with different characteristics: the prefill phase, which is latency-sensitive and scales quadratically with prompt length, and the decode phase, which progresses token by token with undetermined length. Current GPU power governors (for example, NVIDIA default) overlook this asymmetry, treating both phases uniformly. The result is mismatched voltage/frequency settings, leading to suboptimal voltage/frequency configurations, head-of-line blocking, and excessive energy consumption.

We introduce \textsc{GreenLLM}, a service-level objectives (SLO) aware serving framework that minimizes GPU energy by explicitly separating prefill and decode control. At ingress, requests are routed into length‑based queues so short prompts avoid head‑of‑line blocking, tightening TTFT. For prefill, \textsc{GreenLLM} collects short traces on a GPU node, fits compact latency–power models over SM frequency, and solves a queueing‑aware optimization to pick energy‑minimal clocks per class. During decode, a lightweight dual‑loop controller tracks throughput (tokens-per-second) and adjusts frequency with hysteretic, fine‑grained steps to hold tail TBT within target bounds. Across Alibaba and Azure trace replays, \textsc{GreenLLM} achieves up to 34\% reduction in total energy consumption compared to the default DVFS baseline in Alibaba/Azure trace replays, with no loss of throughput and only less than 3.5\% SLO violations increase, demonstrating its effectiveness in the efficient LLM service.



\end{abstract}



\keywords{GPU, LLM, SLO, Energy Optimization, Performance, DVFS}


\maketitle

\input{s1_intro}
\input{s2_background}

\input{s3_proposed}

\input{s4_exp_setup}

\input{s5_results}


\input{s7_stoa}

\section{Conclusion}


In the paper, we have proposed \textbf{GreenLLM}, an adaptive SLO-aware dynamic frequency scaling framework for energy-efficient LLM serving. Our framework couples 1. length-aware routing to mitigate head-of-line effects and preserve TTFT for short prompts, 2. a queueing-aware, model-driven optimizer that selects energy-minimal SM clocks for prefill within latency constraints, and 3. a lightweight, dual-loop token-tracking DVFS controller that holds decode TBT targets while minimizing energy. The approach runs on commodity GPUs, requires no model changes, and integrates cleanly with existing serving stacks.

Across Qwen-14B (Desne) and Qwen-30B (MoE) on real Alibaba/Azure traces, GreenLLM consistently reduces node energy while maintaining throughput and high SLO pass rates, with the largest gains when decode has SLO slack room.
The total energy drops by \textbf{10--34\%} compared with Nvidia's default governor.
Looking ahead, GreenLLM’s principles can extend to larger clusters and next-generation GPUs for better efficiency and sustainability.

\bibliographystyle{ACM-Reference-Format}
\bibliography{references}

\end{document}

%% file: s1_intro.tex
\section{Introduction}
\label{sec:intro}


Generative Large Language Models (LLMs) have experienced explosive adoption in recent years, driving a surge in demand for GPU clusters to support inference workloads~\cite{mckinsey2023genai,kwon2023pagedattention,aminabadi2022deepspeedinference,vllm-github}. To keep up with user requests for chatbots, virtual assistants, and other generative services, enterprises are rapidly scaling up dedicated LLM inference clusters~\cite{aws_inf2_2023,google_tpu_v5e_2023,azure_maia_2024,coburn2025mtia2i,kwon2023pagedattention,aminabadi2022deepspeedinference,azure_openai_chatgpt_2023}. However, a central challenge in this expansion is the substantial power consumption of modern AI accelerators~\cite{yu2023know}. Recent studies estimate that OpenAI’s GPT-4o consumes nearly 1.8 Wh per long query~\cite{jegham2025hungryAI}. At scale, the energy are staggering: with GPT-4o projected to handle about 772 billion queries in 2025, the total electricity demand could reach 391–463 GWh, comparable to the annual consumption of 35,000 U.S. households~\cite{jegham2025hungryAI}.

\begin{figure}[!t]
  \centering
  \includegraphics[width=1.0\linewidth]{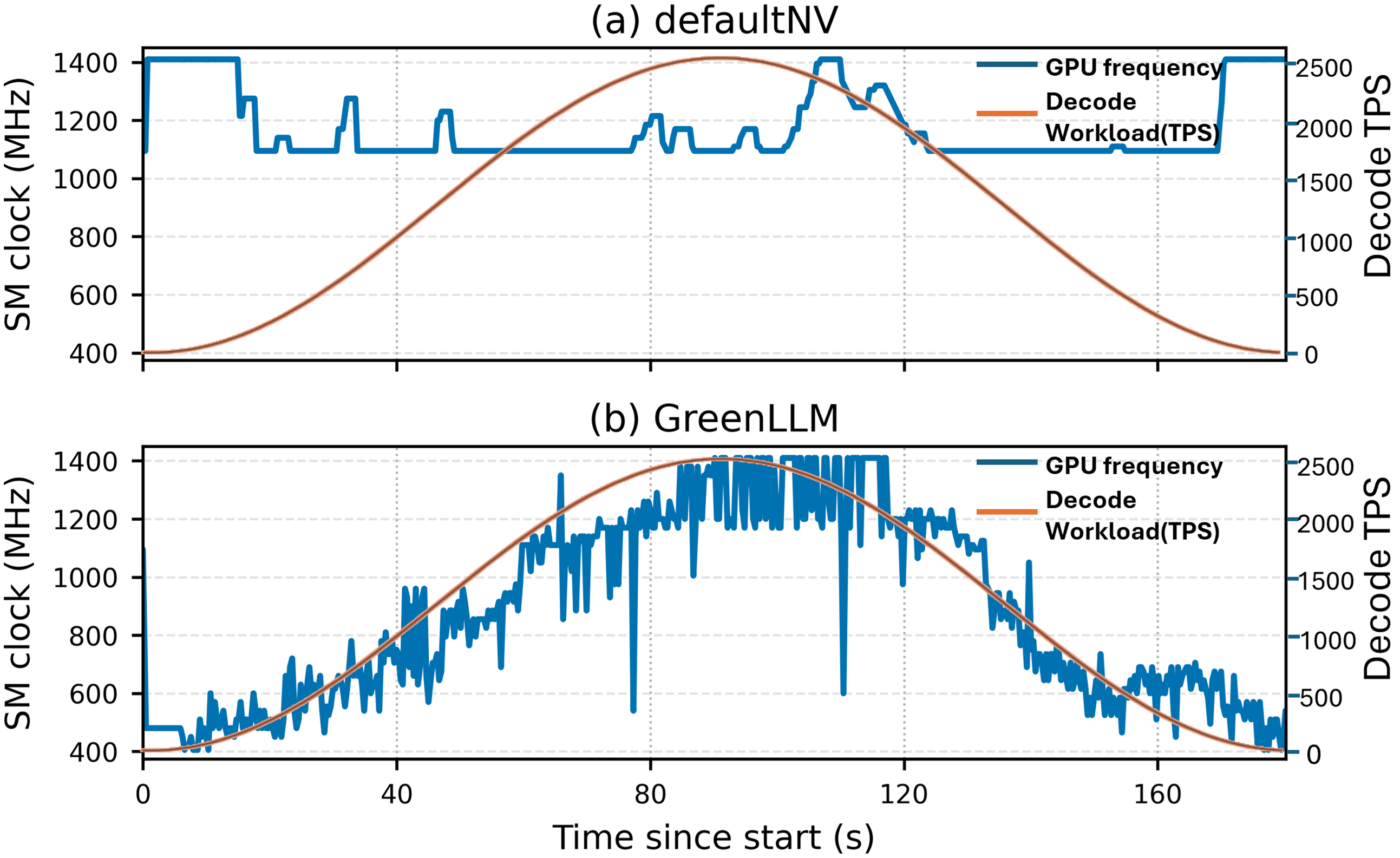}
  \caption{GPU Frequency vs. Decode TPS under defaultNV and GreenLLM}
  \label{fig:TBT_tracing_decode}
\end{figure}

GPUs are the primary workhorses for LLM inference, but they are also the dominant source of power consumption in serving these models. For instance, NVIDIA’s H100 consumes up to 700 W per GPU~\cite{trgdatacenters_h100_power}, with high-density AI racks demanding 30–40 kW~\cite{powerdensities_ai_racks}. 
However, conventional GPU power management and scheduling techniques are often ill-suited for LLM inference. Default GPU frequency scaling governors (e.g., NVIDIA’s) typically drive accelerators at high clock speeds to maximize throughput, with little regard for workload-specific characteristics or latency SLOs~\cite{tang2019impact}. As shown in Fig.~\ref{fig:TBT_tracing_decode}a, under a sinusoidal decode workload, NVIDIA’s default governor (defaultNV) fails to adjust frequency in response to workload intensity. In contrast, Fig.~\ref{fig:TBT_tracing_decode}b illustrates our proposed method, which successfully tracks the workload dynamics.

Similarly, cluster-level power and thermal management mechanisms, such as power capping or throttling, treat all GPU tasks uniformly and usually activate only in emergency conditions, resulting in suboptimal efficiency~\cite{patel2023towards,zhou2022pets,zhang2021flex}. To address these shortcomings, researchers have explored more adaptive strategies, such as fine-grained frequency scaling across different AI operators~\cite{wang2025asplos_dvfs} and SLO-aware methods~\cite{kakolyris2024slo}. Wilkins \textit{et al.}~\cite{wilkins2024offline} develop offline energy models of LLM inference (parameterized by input/output tokens) to guide scheduling across heterogeneous systems, while \emph{throttLL’eM}~\cite{kakolyris2025throttll} predicts the per-query load and dynamically scales GPU frequencies, saving energy under strict latency constraints. However, these approaches operate at coarse granularity (e.g., cluster-level or offline scheduling) or depend on complex forecasting for each query.

Importantly, none explicitly exploits a fundamental property of LLM service: inference follows a two-phase execution pattern, as illustrated in Figure~\ref{fig:llm_pd}, with different characteristics~\cite{agrawal2024taming,patel2024splitwise,zhong2024distserve}, creating new opportunities for fine-grained, workload-specific power management.

In this work, we propose GreenLLM, an energy-aware LLM serving framework that applies phase-specific GPU frequency control to reduce energy consumption under strict SLO constraints. At the prefill phase, GreenLLM first classifies incoming requests into separate queues by prompt length (e.g. short vs. long prompts). This length-based routing isolates short prompts from long ones, preventing head-of-line blocking and significantly improving the time-to-first-token (TTFT) latency for short queries. Then, for each prompt-length class, we profile the LLM’s prefill execution on an NVIDIA DGX-A100 (8× A100 GPUs) server and fit a compact latency–power model as a function of the GPU Streaming Multiprocessor (SM) frequency. Using this model, at runtime, GreenLLM can solve a queueing-aware optimization problem to select the optimal SM frequency to execute prefill phases, maximizing energy efficiency while ensuring incoming requests meet latency SLO.

During the decode phase, GreenLLM employs a lightweight dual-loop feedback controller that dynamically adjusts the GPU frequency in real time. Every 20 ms, the controller measures the current tokens-per-second (TPS) output rate and consults a TPS-to-frequency table (profiled offline) to first narrow down to a coarse-grained frequency band. It then applies fine-grained frequency adjustments in 15 MHz frequency steps (with hysteresis) to precisely maintain the 95th-percentile time-between-tokens (TBT) within the target bound, all while minimizing energy consumption. In essence, GreenLLM uses a fast TPS-feedback loop to throttle down the GPU when full speed is not needed during decoding, and to ramp up when needed to avoid violating latency SLOs (shown as Figure~\ref{fig:TBT_tracing_decode}(b)GreenLLM, detail explained in Sect.~\ref{subsubsec:dynamic_tracing}).

By decoupling the two phases and tailoring the voltage-frequency settings of the GPU to the needs of each phase and each category of prompt length, GreenLLM substantially reduces node-level energy consumption for LLM inference. Our prototype implementation on A100 GPUs demonstrates up to 34\% reduction in total energy consumption compared to NVIDIA’s default DVFS policy on real LLM traffic traces (from Alibaba and Azure), with less than 3.5\% queries missing their latency targets.

%% file: s2_background.tex
\section{Background and Motivation}


To optimize LLM serving, it is necessary to first understand its serving pipeline and how it interacts with modern GPU power management.

\subsection{LLM Inference: Prefill vs. Decode}

\begin{figure}[!t]
  \centering
  \includegraphics[width=1.0\linewidth]{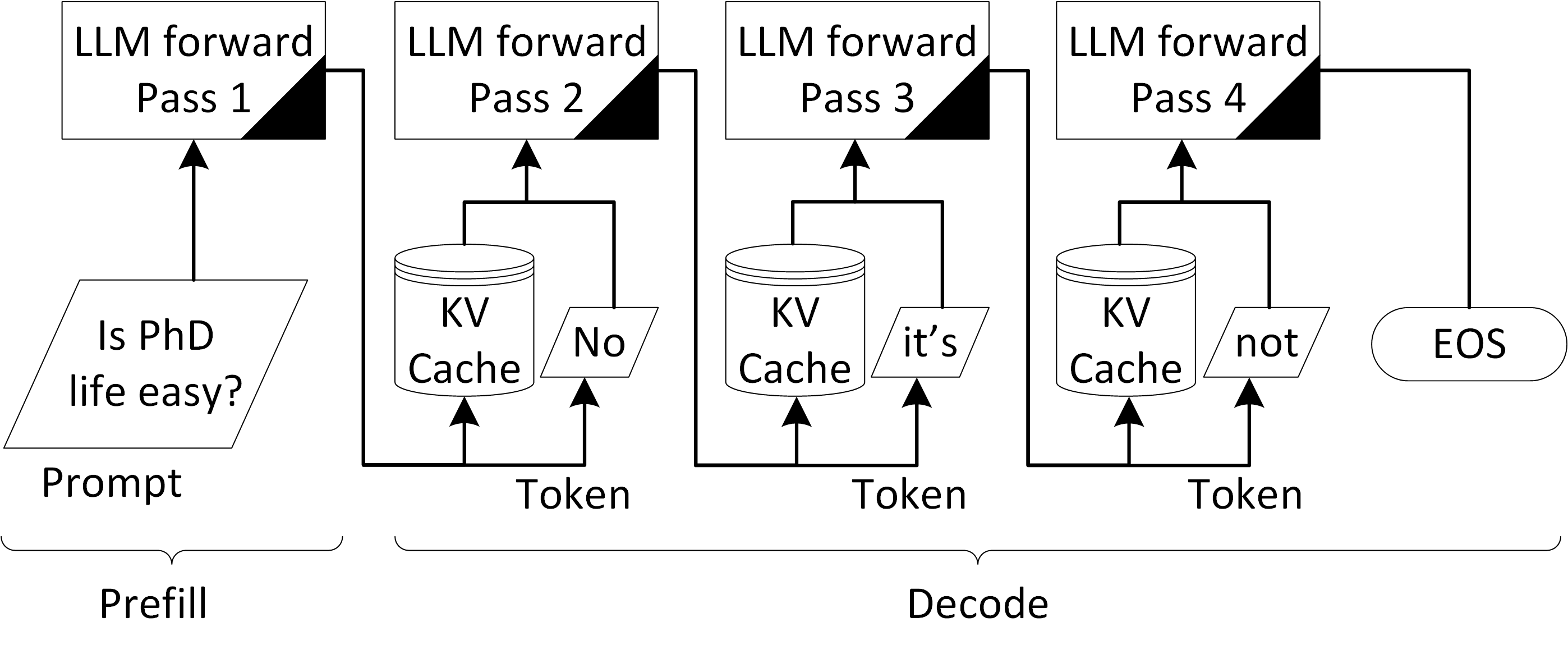}
  \caption{LLM Inference Serving: Prefill and Decode.}
  \label{fig:llm_pd}
\end{figure}

As illustrated in Figure~\ref{fig:llm_pd}, LLM inference comprises two distinct phases with different characteristics~\cite{zhong2024distserve}. When a request arrives, the model first processes the entire input prompt in a prefill phase, then generates output tokens one-by-one in a decode phase. 
Prefill is compute-bound and $\sim O(n^{2})$ in prompt length; it largely determines TTFT.
This phase is typically latency-sensitive, since users await the model’s first response token. In contrast, the decode phase unfolds as an iterative, stepwise generation: the model produces one token at a time, appending it to the context and repeating until completion. The runtime of the decode phase is proportional to the (unpredictable) length of the generated sequence, introduces a streaming aspect, the user gradually receives tokens, and the service often has a target time between tokens (TBT) to keep the output responsive.

These inherent differences mean that the optimal execution strategy can differ between prefill and decode. Prefill benefits from aggressive resource usage to minimize one-time latency, whereas decode might afford more relaxed pacing as long as token stream timing stays smooth. Unfortunately, traditional power managers do not distinguish between these phases~\cite{kakolyris2025throttll,kakolyris2024slo,patel2024characterizing,wang2025asplos_dvfs,wilkins2024offline,stojkovic2025dynamollm,stojkovic2025tapas}, treating an inference query as one monolithic task. This overlooks opportunities to specialize the scheduling and frequency of the GPU for the real needs of each phase.

\subsection{GPU Frequency, Power, and Performance Trade-offs}

Modern GPUs employ dynamic voltage and frequency scaling (DVFS) to balance performance and power consumption. GPU power consumption increases with both supply voltage and operating frequency, following the relation $P \propto V^2 f$~\cite{harris2021digital}. Since voltage scales approximately linearly with frequency, power can be approximated as $P \propto f^3$, providing substantial potential for energy reduction through frequency scaling.

To characterize these relationships for LLM inference, we profile Qwen3-14B on an A100-40GB DGX node using controlled microbenchmarks across varying frequencies and workload intensities (detailed in Section~\ref{sec:setup}).

\begin{figure*}[t]
  \centering
  \begin{subfigure}[t]{0.32\textwidth}
    \centering
    \includegraphics[width=1.0\linewidth]{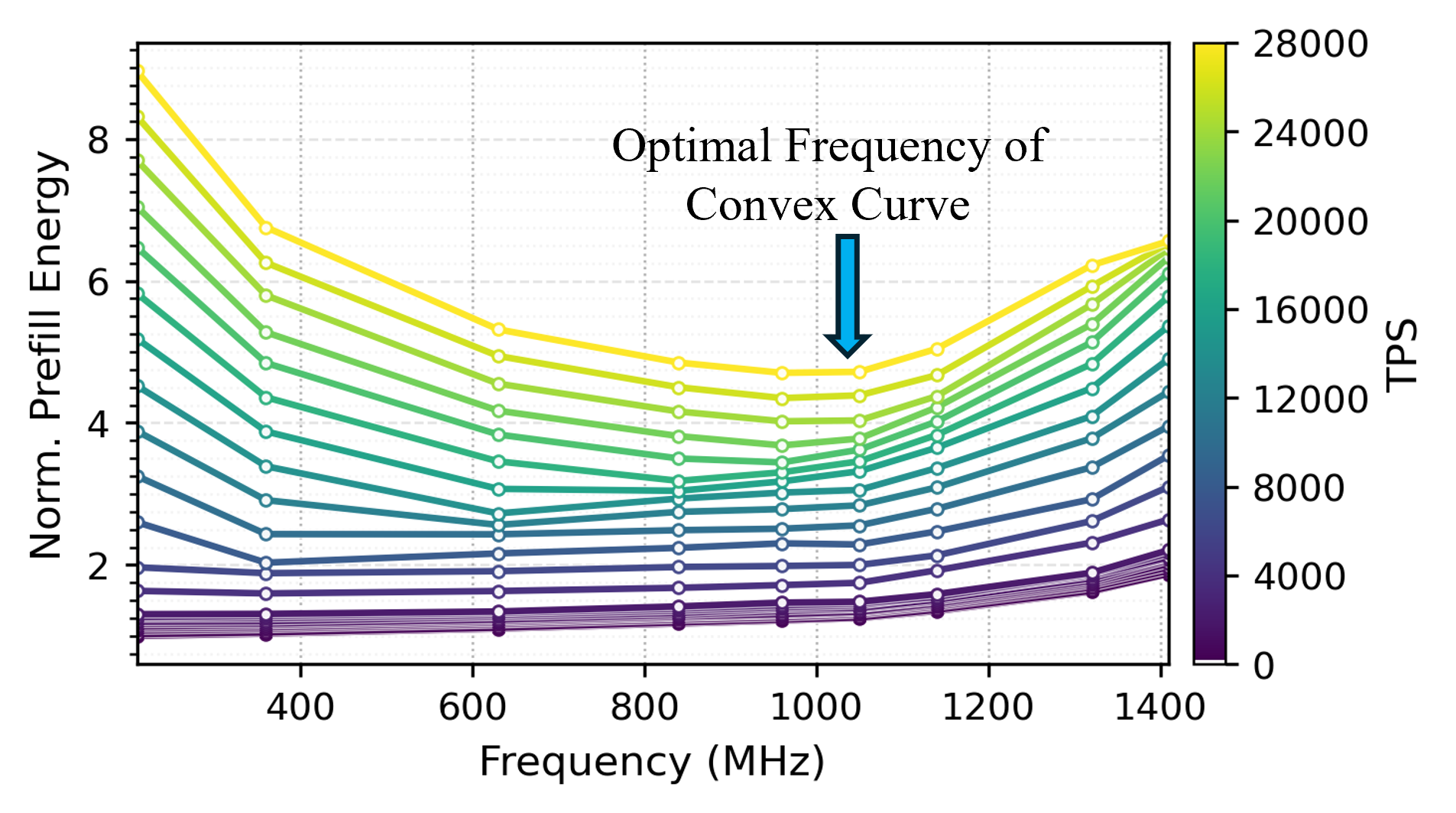}
    \caption{Normalized Prefill Energy vs.\ SM Frequency}
    \label{fig:fig2a}
  \end{subfigure}\hfill
  \begin{subfigure}[t]{0.32\textwidth}
    \centering
    \includegraphics[width=1.0\linewidth]{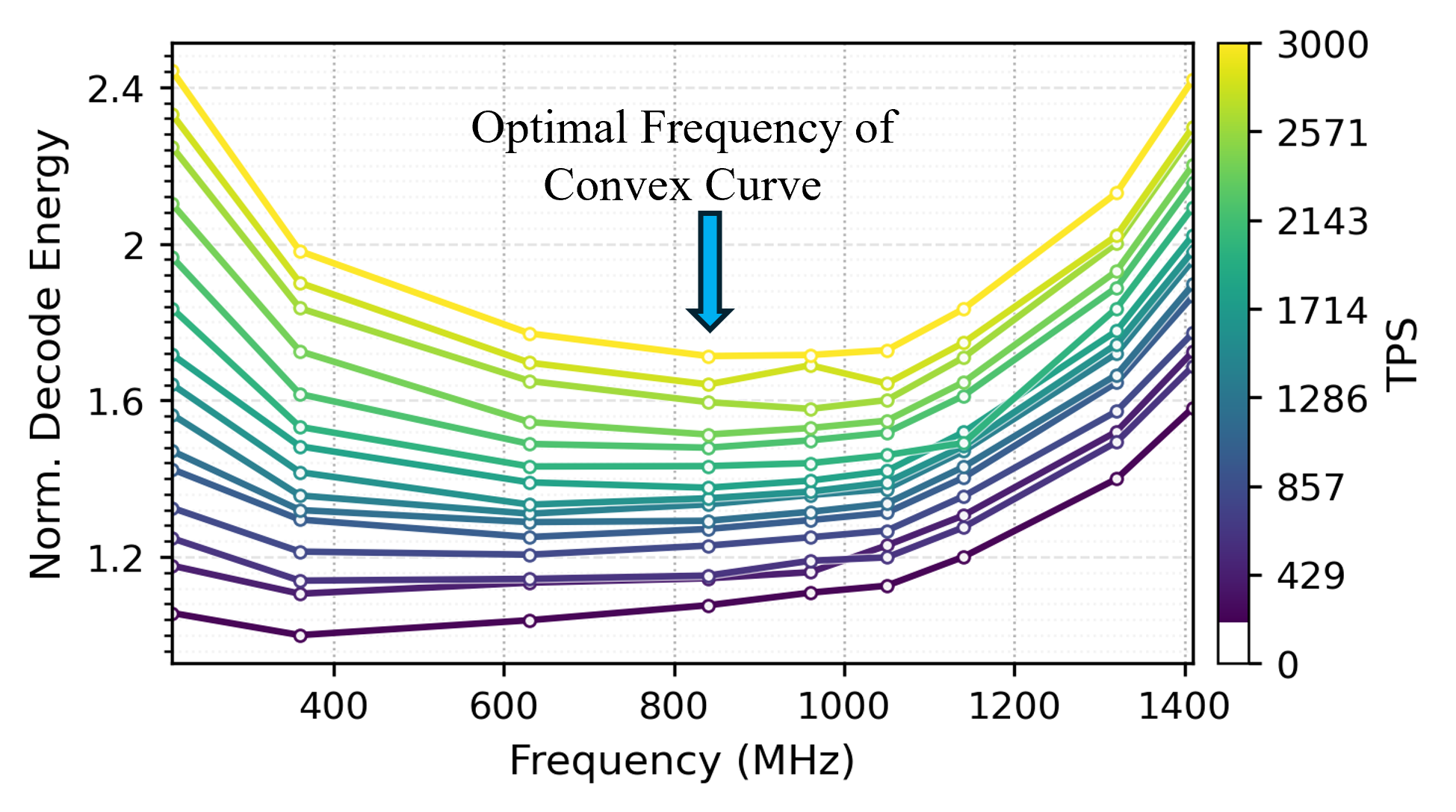}
    \caption{Normalized Decode Energy vs.\ SM Frequency}
    \label{fig:fig2b}
  \end{subfigure}\hfill
  \begin{subfigure}[t]{0.32\textwidth}
    \centering
    \includegraphics[width=1.0\linewidth]{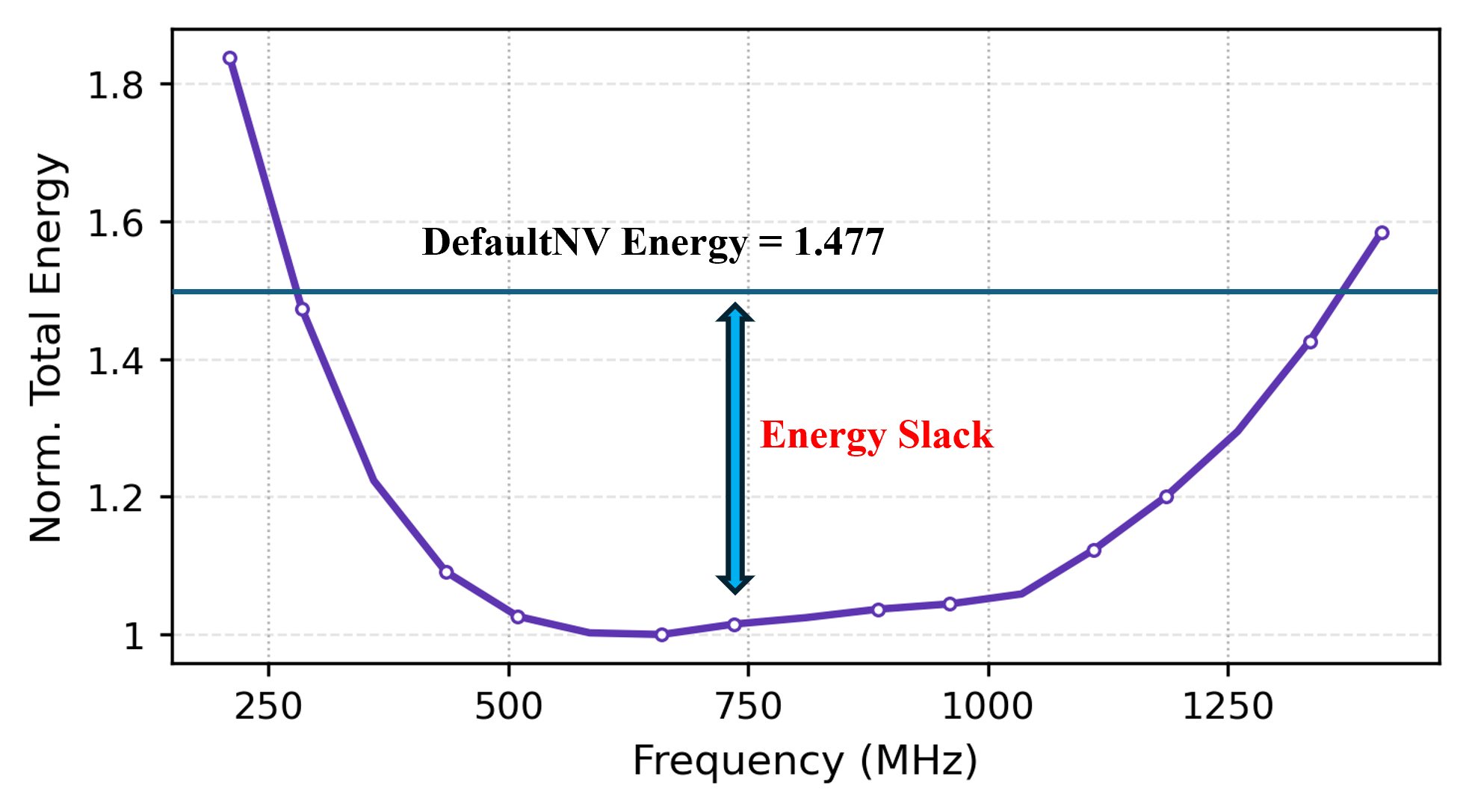}
    \caption{Normalized Total Energy from Practical Trace vs.\ SM Frequency}
    \label{fig:fig2c}
  \end{subfigure}
  \caption{System Energy/Frequency Profiling experiments with different traces}
  \label{fig:fig2}
\end{figure*}

\subsubsection{Profiling with microbenchmark}
\label{subsubsec:microbench}
For a better profiling, we construct two trace-based microbenchmarks. 
Each replays short traces at fixed aggregate TPS targets, while we sweep the GPU \textbf{SM clock} via NVML app-clocks~\cite{nvidia-nvml}; memory clocks are pinned and autoboost is disabled to isolate SM-frequency effects (details described in Sect.~\ref{sec:setup}).

For \textbf{Prefill microbenchmark}, each trace runs prefill and then emits exactly one decoded token to terminate generation. Prompts are length-randomized within 256$-$1024 tokens. We set the TPS range from a 200 TPS to 30000 TPS, and for each TPS level, we vary the SM frequency from lowest 210MHz to highest 1410MHz. 

For \textbf{Decode microbenchmark}, each trace begins with a very short prefill (32 tokens) and then decodes with per-stream generated lengths sampled from $[256,1024]$ tokens. We maintain the target TPS (200–3000) by adjusting concurrency and report the TBT versus the SM frequency. Traces are executed end-to-end (no pre-warmed KV cache).

Using these two microbenchmarks, we sweep SM clocks to quantify phase-specific frequency–power–latency trade-offs; Figures~\ref{fig:fig2a}–\ref{fig:fig2c} report the resulting U-shaped energy profiles and distinct knees for prefill and decode.

\subsubsection{System Profiling}
Figure~\ref{fig:fig2a} shows \emph{normalized prefill energy} ($E/E_{\min}$) versus SM frequency for the prefill stage at different TPS levels. Most of the TTFT curves are convex: performance improves dramatically from low to mid‑range frequencies, then flattens at higher frequencies. 
Energy drops sharply from very low clocks ($\le$400\,MHz) to \mbox{$\sim$0.9--1.0\,GHz}, reaches a broad minimum, and then rises again beyond \mbox{$\sim$1.1--1.2\,GHz}, yielding a clear U-shaped (convex) curve. This shape holds across loads (color-coded by TPS): higher TPS shifts the curves upward but the energy-minimizing frequency remains in a narrow band around \mbox{0.95--1.05\,GHz}, about 70--80\% of the 1.41\,GHz max (check detail explanation and math-metical analysis in Sect.~\ref{subsec:prefill_modeling}). 



\begin{tcolorbox}[takeaway, title=Takeaway \#1]
Prefill Energy shows a convex trend with a broad minimum 
around 0.95–1.05 GHz ($\approx$70–80\% of 1.41 GHz), 
indicating the headroom for frequency-energy optimization.
\end{tcolorbox}

Figure~\ref{fig:fig2b} reports normalized decode energy ($E/E_{\min}$) versus SM frequency. The curves are convex: increasing frequency from low to moderate values reduces energy, but beyond $\sim$1.2\,GHz the energy rises again. Decode is largely \emph{memory-bound} due to intensive key–value (KV) cache reads, so at high clocks GPU SMs spend more time stalled on the HBM/L2/NvLink fabric rather than on arithmetic units~\cite{kakolyris2024slo,prabhu2025vattention,zhao2024alisa}. As a result, the time per token saturates with frequency, while the power grows superlinearly with the clock (roughly $P\!\propto\! fV^2$~\cite{venkatachalam2005power}). The limited latency improvement coupled with higher power yields the observed U-shaped energy curve. Past the knee, decode energy increases more steeply than prefill, which remains more compute-bound and still gains meaningful speedups at higher clocks. Therefore, also, the optimal frequency point for decode worker shows a obvious lower value compared to the prefill worker.

\begin{tcolorbox}[takeaway, title=Takeaway \#2]
The decode stage also shows a convex energy trend, suggesting room for optimization. However, its optimal frequency band is clearly lower than that of prefill, indicating that optimization should begin with a different strategy.
\end{tcolorbox}

Furthermore, Figure~\ref{fig:fig2c} profiles total energy consumption from practical traces (chat traces with 5 queries per second from Alibaba, explained in Sect.~\ref{sec:setup}) under different fixed frequency settings. The energy-frequency curve is also convex: running too slow (left side) prolongs execution (due to the SLO-violate latency) and inflates energy, while running at the highest frequency (right side) also increases energy due to disproportionate power draw, with an optimal sweet spot in between. In our measurements, an intermediate DVFS level minimized energy, for example, capping the GPU clocks around 0.75\,GHz reduced total inference energy by $\sim$47\% compared to the default performance governor (which drives frequencies near peak). The lower optimal frequency compared to Fig.\ref{fig:fig2a}–\ref{fig:fig2b} stems from the trace’s low decode TPS, which shifts the decode knee toward lower frequencies (e.g., Fig.\ref{fig:fig2b} at TPS $\approx$ 200).

\begin{tcolorbox}[takeaway, title=Takeaway \#3]
On practical traces, total energy is convex with a clear minimum, motivating SLO-aware power management.
\end{tcolorbox}

These measurements reveal a consistent U‑shaped relation between energy and frequency across prefill, decode, but with different knees for the two phases (Fig.~\ref{fig:fig2}a-b). Running prefill near a mid‑frequency band minimizes energy while meeting TTFT, whereas decode should track the token stream and stay at a lower clock that just satisfies TBT. Consequently, any single governor is intrinsically suboptimal for LLM serving. This motivates GreenLLM: a dual-stage, SLO‑aware optimization strategy that (i) isolates short and long prompts to protect TTFT, (ii) solves a queueing‑aware optimization to pick prefill clocks per class, and (iii) uses a lightweight TPS/TBT feedback loop to hold decode latency while biasing clocks downward.

%% file: s3_proposed.tex
\section{GreenLLM: SLO-Aware Dual‑Stage LLM energy optimization}
\label{sec:greenLLM}

\begin{figure}[!t]
  \centering
  \includegraphics[width=1.0\linewidth]{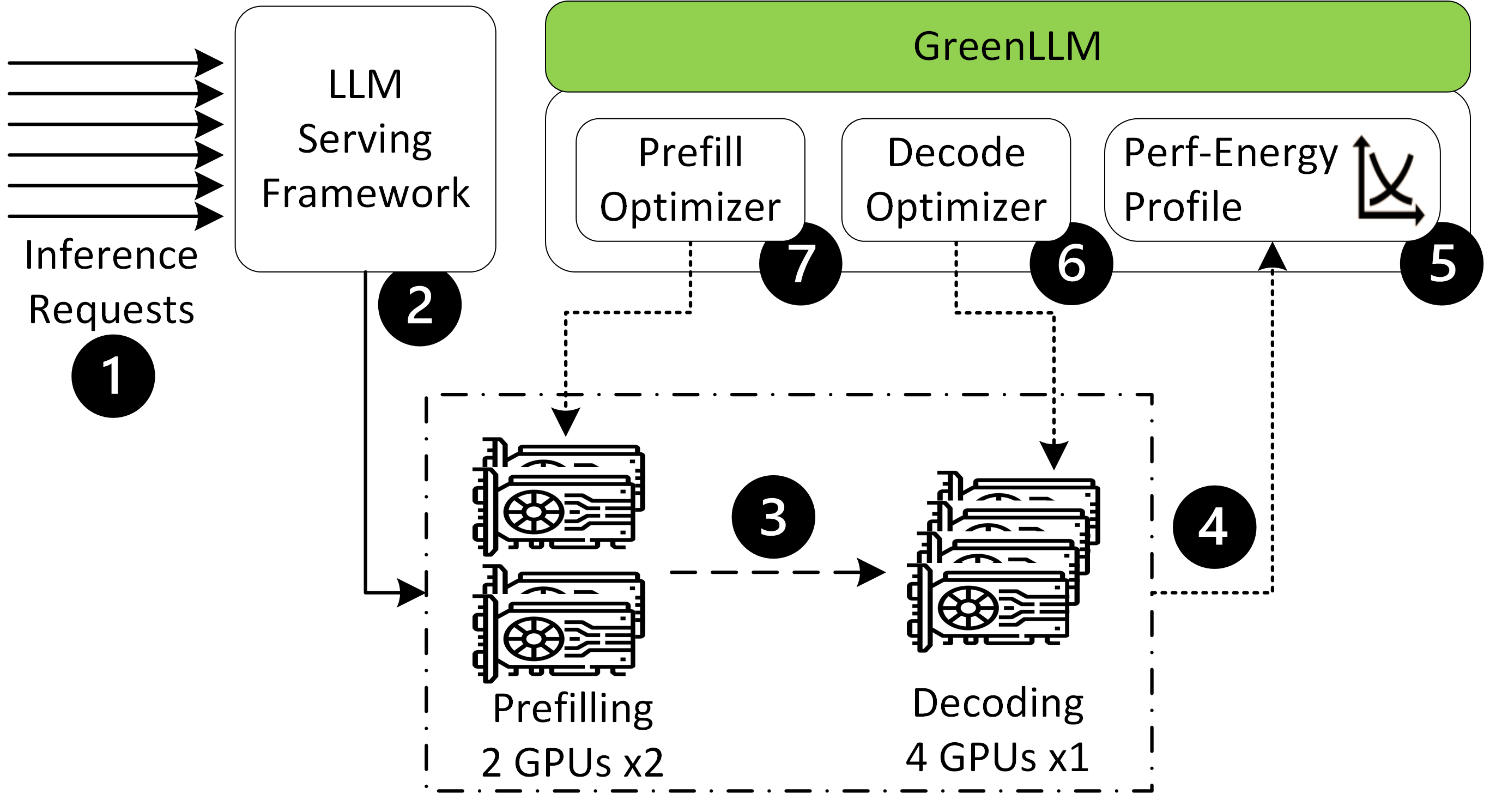}
  \caption{System Overview: Queue-aware prefill optimizer and dual-loop dynamic decode optimizer.}
  \label{fig:greenLLM}
\end{figure}

Figure~\ref{fig:greenLLM} depicts GreenLLM’s control planes. \circlednum{1} Inference requests arrive at the serving framework, which tokenizes and dispatches them to two execution pools. \circlednum{2} A Prefill pool (two workers, each using 2 GPUs) computes the KV-cache. \circlednum{3} Outputs are handed to a Decoding pool (four workers, 1 GPU each) for autoregressive generation. 
\circlednum{4} Per-worker telemetry, including throughput (TPS) and token-level latency SLOs—such as time-to-first-token (TTFT) and time-between-tokens (TBT)—is sampled and streamed to GreenLLM’s \circlednum{5} Perf–Energy Profile block.
The GreenLLM solves a queueing-based problem to select per-pool SM frequencies. GreenLLM issues DVFS updates to the decode \circlednum{6} and prefill \circlednum{7} pools, minimizing energy while meeting SLO constraints.

\subsection{Adaptive Prompt Routing}
\label{subsec:prompt_route}
Mixing short and long prompts in the same serving queue severely hurts latency due to head-of-line (HoL) blocking~\cite{patke2024queue,zheng2023response}. When a lengthy prompt precedes shorter queries, those short requests get stuck waiting, resulting in a heavy long-tail latency distribution where many requests exceed their SLOs due to queueing delays rather than processing time.
To address this issue, we apply a simple length-based routing mechanism that partitions requests by prompt size and sends them to different specialized LLM workers. Instead of waiting for all queries in one queue, we configure $(n-1)$ threshold cut-off points on the prompt length to divide traffic among $n$ workers based on the size of the input (n is 2 in our design). The intuition is straightforward: short prompts go to a worker optimized for speed on small inputs, while extremely long prompts go to a separate worker designed for heavy, long-sequence processing. 
To analytically explain the impact from mixed and separate prefill, we classify the prompts into two classes short-medium prompts (SM) and the long prompts (L).
For example, with two prefill workers:
\begin{itemize}[leftmargin=*,labelsep=0.4em,itemsep=0pt,parsep=0pt,topsep=2pt]
    \item \textbf{Short-Context Model:} Optimized for low-latency processing of SM inputs (up to \textit{approximately 1024 tokens}). This worker can rapidly handle the common, shorter queries without being bogged down by extremely long tasks.
    \item \textbf{Long-Context Model:} Equipped to support much larger prompts and optimized for L-sequence processing. This worker handles the infrequent but expensive long queries on a separate track, so they don’t block the short ones.
\end{itemize}

\begin{figure}[!t]
  \centering
  \includegraphics[width=0.75\linewidth]{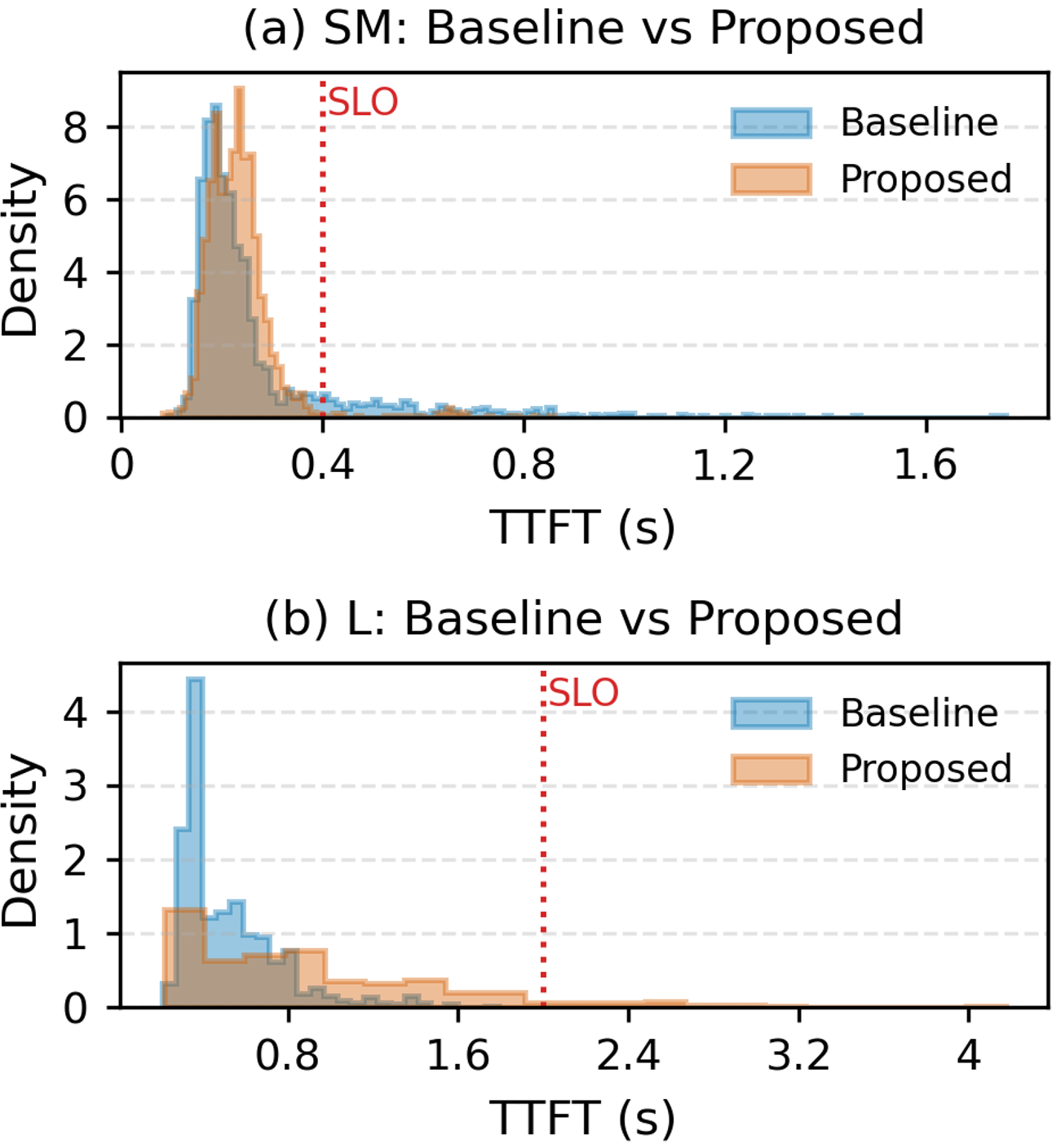}
  \caption{TTFT distribution before routing (a) and after length-based routing (b). Routing separates workloads by prompt length, lowering latency for short/medium queries while keeping long queries within their SLO.}
  \label{fig:ttft_dist}
\end{figure}

This simple length-based partitioning avoids the need for complex scheduling logic. By isolating long requests away from the main short-request queue, we eliminate most HoL blocking between dissimilar query lengths.
The Orange histogram in Figure~\ref{fig:ttft_dist}(a) shows the profiling experiment running the Alibaba's chat trace with 8 queries-per-second on our system (check Sect.~\ref{sec:setup} for detail) with default method and the adaptive prompt routing.
This length-aware routing dramatically reduces the SLO violations. The bulk of requests, which tend to have short or medium prompts, now achieve consistently low SLO violations because these short queries are no longer delayed by rare long requests in the same queue, so their tail latency drops significantly. Meanwhile, the long prompts are still processed in a reasonable time on their dedicated worker, but crucially their slower processing no longer interferes with the vast majority of traffic.
By aligning each request to an appropriate execution path, we better satisfy divergent latency objectives across mixed workloads: fast service for short prompts and efficient handling of long prompts, with minimal mutual impact. In total, the percentage of the requests meeting the SLO increase sharply from 89.9\% to 96.4\%.



\subsection{Prefill Modeling and Optimization}
\label{subsec:prefill_modeling}
\begin{figure}[!t]
  \centering
  \includegraphics[width=1.0\linewidth]{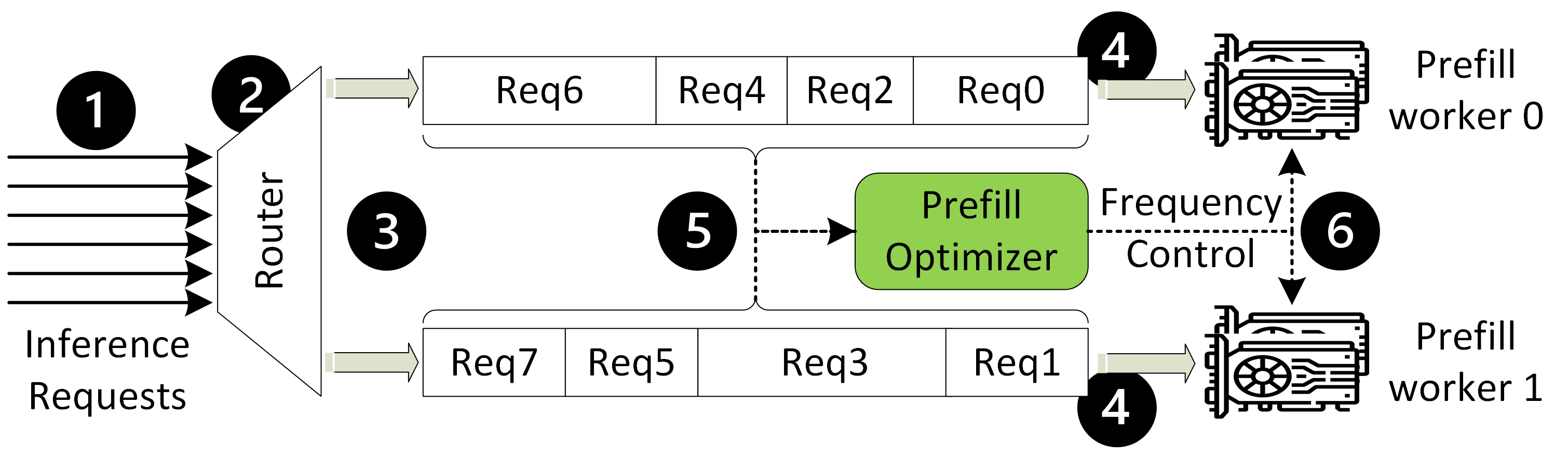}
  \caption{Prefill queue and control: prefill optimizer sets SM frequency to meet TTFT with lower energy}
  \label{fig:prefillQueue}
\end{figure}

As mentioned in Figure~\ref{fig:prefillQueue} and Section~\ref{subsec:prompt_route}, the queue is an important source of delay. We treat the observed queueing as direct information to start the optimization for the prefill stage.
To act on this signal, we need a service-time model that predicts both latency and energy as a function of prompt length and GPU frequency.

To do that, we first profile the serving stack across a range of prompt lengths on a reference SM clock \(f_{\mathrm{ref}}\) (typically the maximum clock). The prefill pass of a decoder-only Transformer has two dominant components per layer: (i) linear terms from QKV/output projections and the FFN, and (ii) a quadratic term from causal attention. 
The prefill FLOPs per layer can be summarized as
\begin{equation}
\boxed{\mathrm{FLOPs}_{\text{prefill}/\ell}(B,n)\;=\;A\,n \;+\; C\,n^{2},}
\label{eq:prefill_flops}
\end{equation}
with
\[
\begin{aligned}
A &= 8B\,d_{\mathrm{model}}^{2} + 4B\,d_{\mathrm{model}}\,d_{\mathrm{ff}},\\
C &= 4\alpha B H_q d_k \;\approx\; 4\alpha B d_{\mathrm{model}}
   \quad \text{(since } H_q d_k = d_{\mathrm{model}}\text{).}
\end{aligned}
\]
\emph{Interpretation:} the linear term \(A n\) comes from QKV/output projections and the FFN; the quadratic term \(C n^{2}\) comes from causal attention~\cite{vaswani2017attention}. Here \(d_{\mathrm{model}}\) is the hidden size, \(d_{\mathrm{ff}}\) the FFN width, \(H_q\) the number of query heads, \(d_k\) the head dimension (typically \(H_q d_k = d_{\mathrm{model}}\)), and \(\alpha \approx \tfrac{1}{2}\) if the kernel computes only the causal triangle; otherwise \(\alpha \approx 1\).
Eq.~\eqref{eq:prefill_flops} makes explicit that projections+FFN scale as \(O(n)\) while attention scales as \(O(n^2)\).

To model the prefill-stage latency, we simplify Eq.~\eqref{eq:prefill_flops} into an interpretable quadratic in input length and profile the system at \(f_{\mathrm{ref}}\) using a sweep of prompt lengths. Let \(L_k\) be the tokens of job \(k\). We fit
\begin{equation}
t_k^{\mathrm{ref}} \;\approx\; a\,L_k^{2} + b\,L_k + c,
\label{eq:prefill_poly}
\end{equation}
where \(a\) captures the attention cost, \(b\) captures projections+FFN, and \(c\) absorbs fixed overheads (e.g., tokenization and launches).

\begin{figure}[!t]
  \centering
  \includegraphics[width=0.85\linewidth]{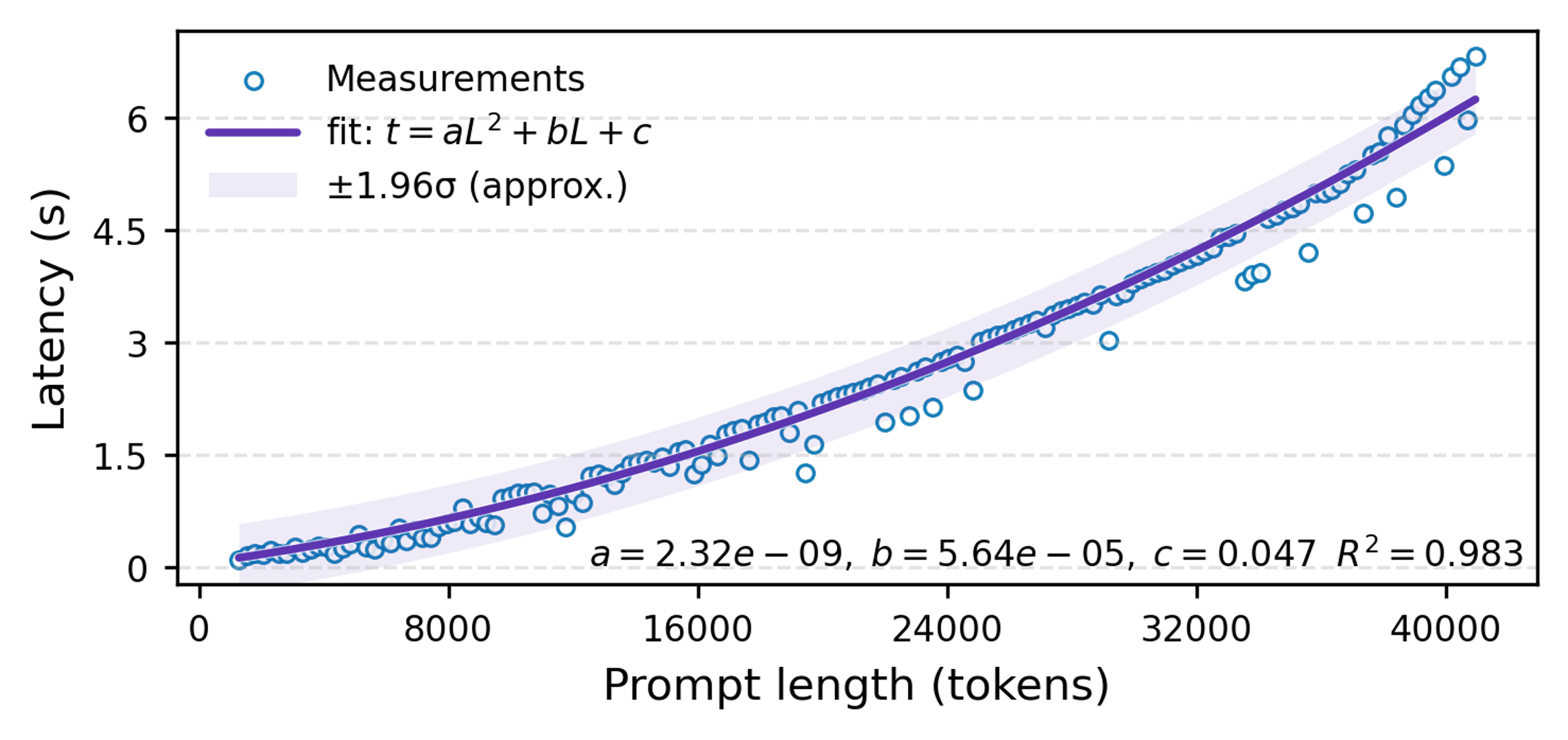}
  \caption{Prefill latency vs prompt length for Qwen3‑14B; quadratic fit t = aL² + bL + c to measurements}
  \label{fig:tps-to-freq}
\end{figure}

With real measurements of Qwen-14B, we therefore fit a second-order polynomial to the measured prefill latencies as a function of prompt length, as shown in Figure~\ref{fig:tps-to-freq}.

We then incorporate the effect of DVFS by modeling latency is inversely proportional to frequency (a reasonable first-order assumption since lower clock speeds linearly slow down compute-bound workloads). Thus, for a general frequency \(f\), the prefill latency for job \(k\) is
\begin{equation}
t_k(f) \;=\; t_k^{\mathrm{ref}} \cdot \frac{f_{\mathrm{ref}}}{f}.
\label{eq:dvfs_scaling}
\end{equation}
In practice, when \(t_k^{\mathrm{ref}}\) is not directly measured for a new \(L_k\), we use the fitted model in \eqref{eq:prefill_poly} as a surrogate.
Using this per-job model, we can predict the total prefill \emph{busy time} for a set of jobs as a function of frequency. Suppose that a batch of jobs (or all pending requests in a scheduling interval) is indexed by \(k=1,2,\dots,N\). The total busy time (cumulative time the GPU is actively running prefill computation for all these jobs) at frequency \(f\) is
\begin{equation}
\mathrm{busy}(f) \;=\; \sum_{k=1}^{N} t_k(f).
\end{equation}
Substituting \eqref{eq:dvfs_scaling} gives
\begin{align}
\mathrm{busy}(f)
  \;&=\; \sum_{k=1}^{N} \left( t_k^{\mathrm{ref}} \cdot \frac{f_{\mathrm{ref}}}{f} \right)
   \;=\; \frac{f_{\mathrm{ref}}}{f} \sum_{k=1}^{N} t_k^{\mathrm{ref}}
   \;=\; \frac{f_{\mathrm{ref}}}{f}\, T_{\mathrm{ref}},
   \label{eq:busy_f}
\end{align}
where \(T_{\mathrm{ref}} \triangleq \sum_{k=1}^{N} t_k^{\mathrm{ref}}\) denotes the total prefill execution time for these jobs at the reference frequency.

In our SLO-aware setting, we have a latency target \(D\) (e.g., the maximum allowable time to finish all prefills, derived from the SLO). To meet the SLO, the chosen frequency must satisfy the requisite:
\begin{equation}
\mathrm{busy}(f) \;\le\; D,
\end{equation}
i.e., the total prefill work is completed by the deadline. Higher frequencies reduce busy time but consume more power, while lower frequencies save energy but increase runtime; our goal is to find the optimal balance with the following power model.

\begin{figure}[!t]
  \centering
  \includegraphics[width=0.85\linewidth]{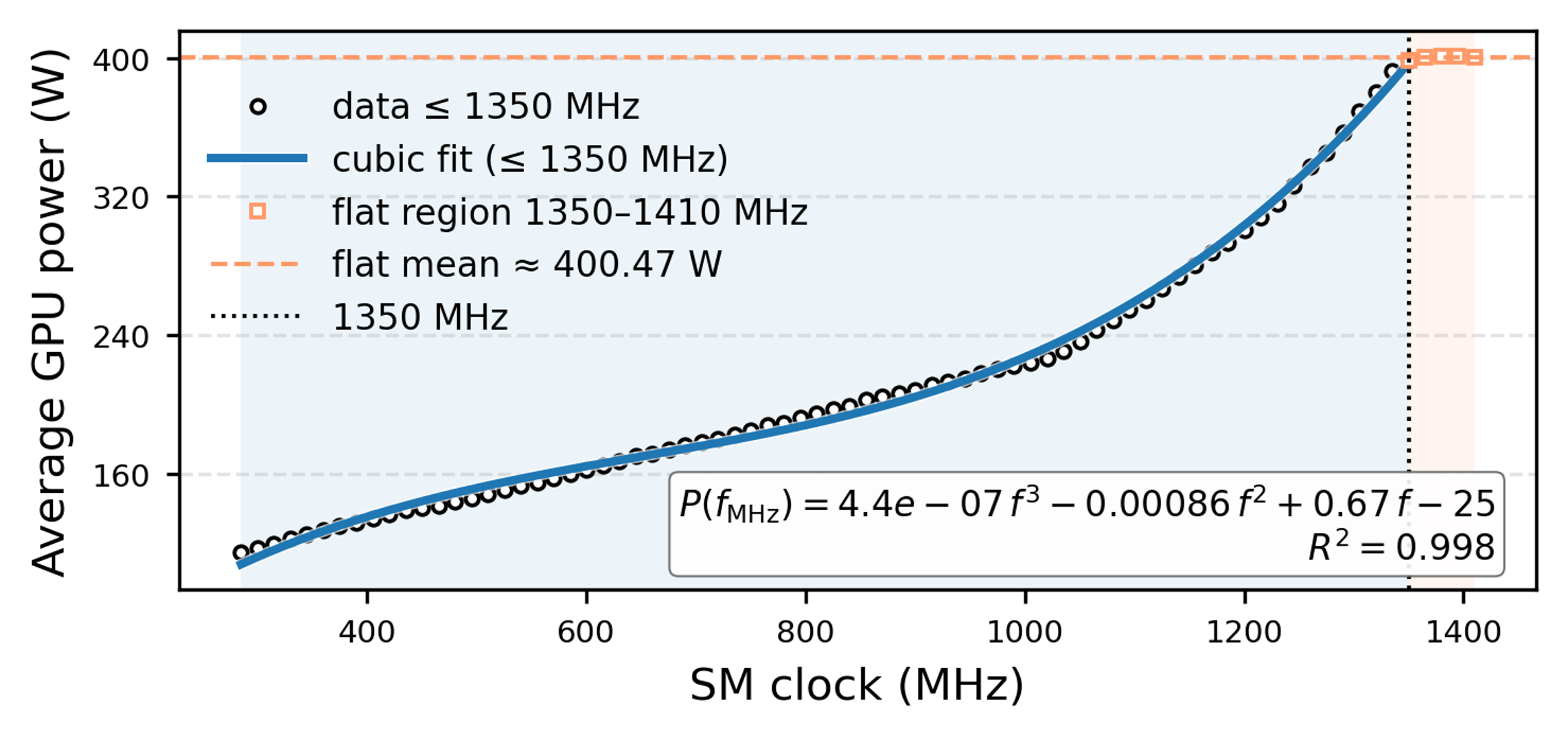}
  \caption{Power modeling with Qwen3-14B inference prefill with varying frequency; measured curve with cubic fit P(f) capturing DVFS scaling.}
  \label{fig:energyfreq}
\end{figure}

We model the power draw and energy consumption of the GPU as functions of frequency. From system profiling, the \emph{active} power (while executing prefill) increases superlinearly with frequency, consistent with CMOS DVFS where dynamic power grows roughly cubically with frequency~\cite{venkatachalam2005power} (via joint voltage--frequency scaling). We thus fit a cubic polynomial to active power over frequency:
\begin{equation}
\begin{aligned}
P(f) \; &=\; k_{3} f^{3} \;+\; k_{2} f^{2} \;+\; k_{1} f \;+\; k_{0}, \\
P_{\text{idle}} \; &=\; P_{0}.
\end{aligned}
\label{eq:power_cubic}
\end{equation}

where $k_{3},k_{2},k_{1},k_{0}$ are regression coefficients.
When the GPU is actively running prefill at frequency $f$, we use $P(f)$ as the instantaneous power draw. $P_{\mathrm{idle}}$ is the background power when the GPU is powered but idle. Note $P_{0} \neq k_{0}$. To build the model, we drive the prefill tier with fixed-length prompts (e.g., 1{,}024 tokens) at a high request rate (40~QPS) to saturate the SMs, then sweep the SM clock and record GPU power. Figure~\ref{fig:energyfreq} reports the measured power–frequency curve. We model average power with a low-order polynomial plus a frequency-independent baseline flat.

Consider an SLO interval of length $D$. Let $\mathrm{busy}(f)$ denote the total prefill busy time (sum of per-job runtimes) at frequency $f$. The \emph{active} energy is
\begin{equation}
E_{\mathrm{active}}(f) \;=\; P(f)\,\mathrm{busy}(f),
\label{eq:active_energy}
\end{equation}
assuming the GPU draws $P(f)$ whenever it is busy. If $\mathrm{busy}(f) \le D$, the remaining time $D-\mathrm{busy}(f)$ is idle, yielding \emph{idle} energy
\begin{equation}
E_{\mathrm{idle}}(f) \;=\; P_{\mathrm{idle}}\,[\,D - \mathrm{busy}(f)\,],
\qquad \text{ when } \mathrm{busy}(f)\le D,
\label{eq:idle_energy}
\end{equation}
otherwise the SLO is violated and such $f$ is infeasible. The total energy within the SLO window is
\begin{equation}
E_{\mathrm{total}}(f) \;=\; E_{\mathrm{active}}(f) \;+\; E_{\mathrm{idle}}(f).
\label{eq:total_energy}
\end{equation}

Using the prefill busy-time model $\mathrm{busy}(f)=\dfrac{f_{\mathrm{ref}}}{f}\,T_{\mathrm{ref}}$ with
\begin{equation}
T_{\mathrm{ref}} \;\triangleq\; \sum_{k=1}^{N} t_k^{\mathrm{ref}} \;\approx\; \sum_{k=1}^{N} \bigl(a L_k^{2} + b L_k + c\bigr),
\label{eq:Tref_def}
\end{equation}
we obtain
\begin{equation}
\resizebox{\columnwidth}{!}{$
\begin{aligned}
E_{\mathrm{total}}(f)
&= (k_{3} f^{3} + k_{2} f^{2} + k_{1} f + k_{0})\,\frac{f_{\mathrm{ref}}}{f}\,T_{\mathrm{ref}}
  + P_{\mathrm{idle}}\!\left[D - \frac{f_{\mathrm{ref}}}{f}\,T_{\mathrm{ref}}\right] \\[2pt]
&= f_{\mathrm{ref}} T_{\mathrm{ref}} \!\left( k_{3} f^{2} + k_{2} f + k_{1} + \frac{k_{0}}{f} \right)
  + P_{\mathrm{idle}}\!\left[ D - \frac{f_{\mathrm{ref}}}{f}\,T_{\mathrm{ref}} \right].
\end{aligned}
$}
\label{eq:etotal_closed_form}
\end{equation}

Because \(E_{\mathrm{total}}(f)\) in~\eqref{eq:etotal_closed_form} is non-monotonic, at runtime, GreenLLM selects the clock by solving a optimization problem under the constrain:
\[
\min_{f\in[f_{\min},\,f_{\max}]} E_{\mathrm{total}}(f)\quad
\text{s.t.}\;\mathrm{busy}(f)\le D.
\]
At runtime, the \emph{Queue Optimizer} solve the optimization problem dynamically and applies the best feasible frequencies on workers to achieve the optimal energy consumption under the SLO constrains.

\subsection{Decode‑Stage optimization}
\label{subsec:decode-dvfs}

\begin{figure}[!t]
  \centering
  \includegraphics[width=0.9\linewidth]{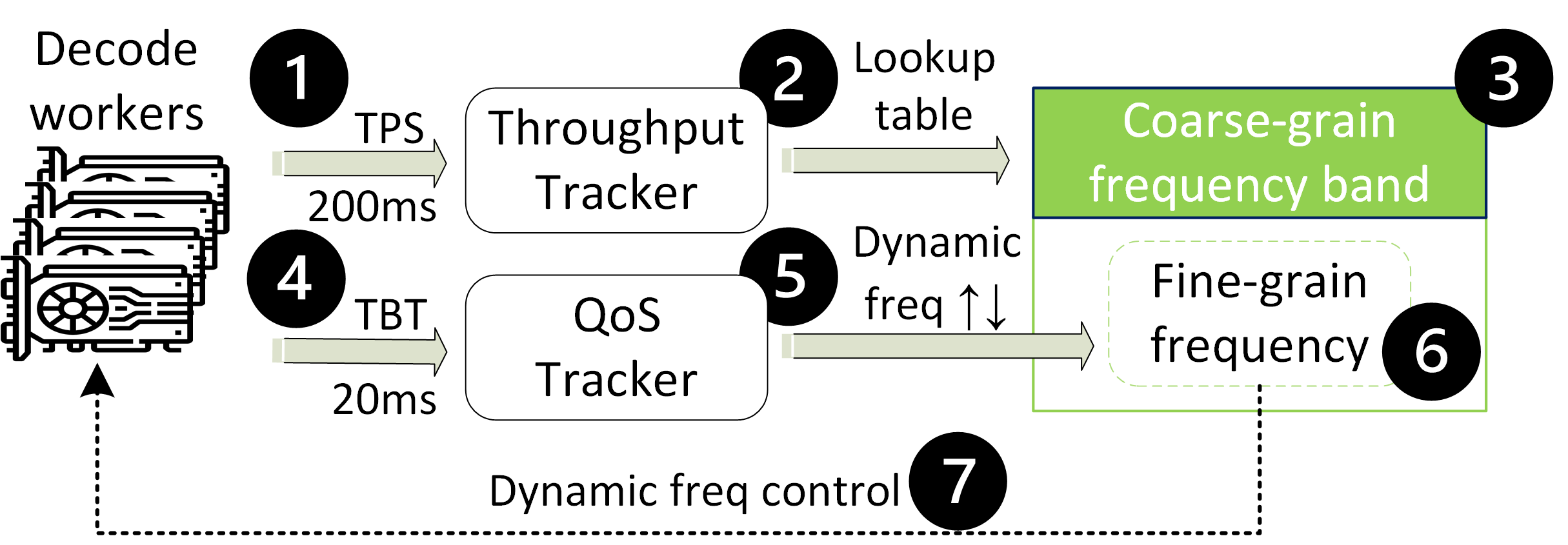}
  \caption{Decode control: TPS determines coarse frequency‑band and fine frequency adjustment with hysteresis to meet P95 TBT.}
  \label{fig:decodecontrol}
\end{figure}

To minimize energy consumption during autoregressive decoding without degrading user-visible performance,
we design a decode stage controller that operates in two coordinated loops, as shown in Figure~\ref{fig:decodecontrol}. We will introduce these two loops in the following subsections. Note that all decisions are made outside the GPU execution path by an asynchronous process that stays outside the critical path.


%

\subsubsection{Coarse-Grain Frequency Band Selection}

The goal of the coarse-grain loop is to estimate decoding load and swiftly select a suitable GPU frequency range that maintains TBT targets with minimal energy consumption.

We track TPS using a sliding window of the past 200\,ms of emitted tokens. This window smooths out momentary bursts and provides a stable estimate of overall decode throughput. To determine optimal frequencies, we conduct an offline profiling sweep using microbenchmark with varying TPS value from 200 to 3000TPs (Sect.~\ref{subsubsec:microbench} and Figure~\ref{fig:fig2b}) across a range of GPU SM clocks. For each TPS bucket, we identify the lowest frequency that satisfies two criteria, namely, maintains P95 TBT below 100\,ms and minimizes energy per token (i.e., power/TPS). These values form a static lookup table: \emph{TPS bucket} $\rightarrow$ \emph{optimal frequency}. At runtime, the controller:
\begin{enumerate}
  \item Map the current TPS to its corresponding bucket, as shown in Figure~\ref{fig:decodecontrol} \circlednum{1}, \circlednum{2}.
  \item Select the frequency band as the triplet: the optimal frequency plus its two neighbors (e.g., $[f_{\mathrm{lo}}, f_{\mathrm{mid}}, f_{\mathrm{hi}}]$)(\circlednum{3}).
  \item Apply hysteresis: the band is updated only if the TPS remains in the new bucket for at least three consecutive 200\,ms intervals. This mechanism balances reactivity with stability, preventing unnecessary frequency jitter.
\end{enumerate}

\subsubsection{Fine-Grain TBT Tracker and Frequency Control}

The fine-grain loop makes fine adjustments to GPU frequency every 20\,ms based on real-time token latency observations(Figure~\ref{fig:decodecontrol} \circlednum{4}). Its role is to track the target TBT (100\,ms P95) and conserve energy by reducing the frequency whenever possible. For every 20\,ms, we compute the P95 TBT in a sliding window of recent tokens. Let
\begin{equation}
  \mathrm{TBT}_{\mathrm{margin}} \;=\; \frac{\mathrm{P95}\ \mathrm{TBT}}{T_{\mathrm{SLO}}}, \qquad
  T_{\mathrm{SLO}} = 100\,\mathrm{ms}.
\end{equation}
The update rule is(Figure~\ref{fig:decodecontrol}(\circlednum{5}, \circlednum{6}):
\begin{itemize}[leftmargin=*,labelsep=0.4em,itemsep=0pt,parsep=0pt,topsep=2pt]
  \item If $\mathrm{TBT}_{\mathrm{margin}} > 1.0$: increase GPU frequency by 15\,MHz (up to the band upper bound).
  \item If $\mathrm{TBT}_{\mathrm{margin}} < 0.65$: decrease frequency by 15\,MHz (not below the band lower bound).
  \item Else: hold the frequency.
\end{itemize}

Each fine-grain adjustment is rate-limited to 15-30\,MHz per control tick to preserve stability. The set point is constrained to the frequency band selected by the coarse-grain loop (Figure~\ref{fig:decodecontrol} \circlednum{3}), based on the current TPS, and its two neighboring bands. The coarse- and fine-grain loops operate in concert: the coarse loop chooses the band, and the fine loop determines the optimal frequency within it.

\subsubsection{TBT Tracker and Coarse-Grain Frequency Band Update}

To address performance drift and workload variability, we update the coarse lookup table using feedback from the fine-grain controller. Every 6 seconds, we sample adjusted frequency events and compute TBT across the window. We apply adaptation logic to shift frequency bands upward or downward when sustained bias is detected (>80\% of adjustments exceed band bounds). This dual-loop mechanism enables fine-grained, SLO-aware control that responds to decoding dynamics in real time.

%% file: s4_exp_setup.tex
\section{Experimental Setup}
\label{sec:setup}

\subsection{Hardware and Software Setup}
\label{subsec:hard_software}
As shown in Table~\ref{tab:hw-sw}, all experiments run on a NVIDIA DGX-A100 node, using NVIDIA Dynamo as deployment framework.

\begin{table}[t]
\centering
\caption{Hardware and software configuration.}
\label{tab:hw-sw}
\begin{tabular}{p{0.30\linewidth} p{0.63\linewidth}}
\toprule
\textbf{Component} & \textbf{Setting} \\
\midrule
Node            & NVIDIA DGX--A100 (1 node)~\cite{nvidia-a100-arch-wp} \\
GPUs            & 8$\times$ A100--SXM4 40\,GB; intra-node NVLink/NVSwitch \\
CPU             & AMD EPYC 7302, 16 cores @ 3.0\,GHz~\cite{amd-epyc-7002-datasheet} \\
Framework       & Dynamo v0.3.1~\cite{nvidia-dynamo-dev} \\
Inference kernel & NVIDIA TensorRT / TensorRT-LLM~\cite{nvidia-tensorrt-doc} \\
DVFS control    & NVML application clocks (SM); memory clocks pinned~\cite{nvidia-nvml,nvidia-smi-man} \\
\bottomrule
\end{tabular}
\end{table}

\subsection{Models and Inference Traces}
\label{subsec:models_traces}
\subsubsection{Models}

We evaluate using a dense model (Qwen3-14B) and a mixture-of-experts model (Qwen3-30B-MoE)~\cite{qwen3technicalreport,qwen3-14b,qwen3-30b-a3b-instruct-2507}, shown in Table~\ref{tab:models}. The system uses two execution pools: a Prefill pool (2 workers, 2 GPUs each) and a Decode pool (4 workers, 1 GPU each).

\begin{table}[t]
\centering
\caption{Models: Qwen3‑14B and Qwen3‑30B, datatype: BF16.}
\label{tab:models}
\begin{tabular}{@{}l l l c c l c l@{}}
\toprule
\textbf{Model} & \textbf{Type} &
\shortstack[l]{\textbf{Params}\\\textbf{(total / active)}} &
\textbf{Layers} & \shortstack[l]{\textbf{Experts}\\\textbf{(active)}} \\
\midrule
Qwen3--14B & Dense & 14.8B / 14.8B & 40 & --- \\
Qwen3--30B & MoE & 30.5B / 3.3B & 48 & 128 (8) \\
\bottomrule
\end{tabular}
\par\noindent\scriptsize\emph{Note:} “Active” parameters are used per token (for dense models, active = total).
\end{table}

\subsubsection{Traces and Evaluation Methods}
We evaluate using: (i) phase-specific microbenchmarks that isolate prefill and decode performance across frequency ranges, and (ii) production traces from Alibaba ServeGen~\cite{servegen} (QPS: 1,3,5,8,10) and Azure 2024~\cite{azure-llm-inference-dataset-2024} (downsampled to 1/8, 1/5, 1/4 rates of its original rate to match single-node capacity (the original targets a GPU cluster) while preserving the inter-arrival structure).

We compare three configurations: \textbf{DefaultNV} (NVIDIA default), \textbf{PrefillSplit} (length-based routing only), and \textbf{GreenLLM} (length-based routing + prefill/decode optimizer). SLOs target TTFT $<400$ms for Short/Medium prompts, $<2$s for Long, and P95 TBT $\le100$ms during decode by following Azure targets~\cite{stojkovic2025dynamollm}.

%% file: s5_results.tex
\section{Results}

We evaluated GreenLLM’s energy–performance tradeoffs through a set of targeted experiments. Our evaluation follows three steps:

(1) Phase-level profiling: We use microbenchmarks to measure the two main inference stages separately: prompt processing (prefill) and token generation (decode). This step clearly demonstrates the different principles, strategies, and benefits for different phases of LLM serving.

(2) End-to-end validation: We then run complete workloads to check whether GreenLLM can meet service-level objectives (SLOs) under realistic conditions at the system level.

(3) Margin analysis: Furthermore, we study how latency slack affects the balance between energy use and SLO compliance, showing the adaptability of our design across a wide range of scenarios.


\subsection{Phase-level profiling}

\begin{figure}[!t]
  \centering
  \includegraphics[width=0.85\linewidth]{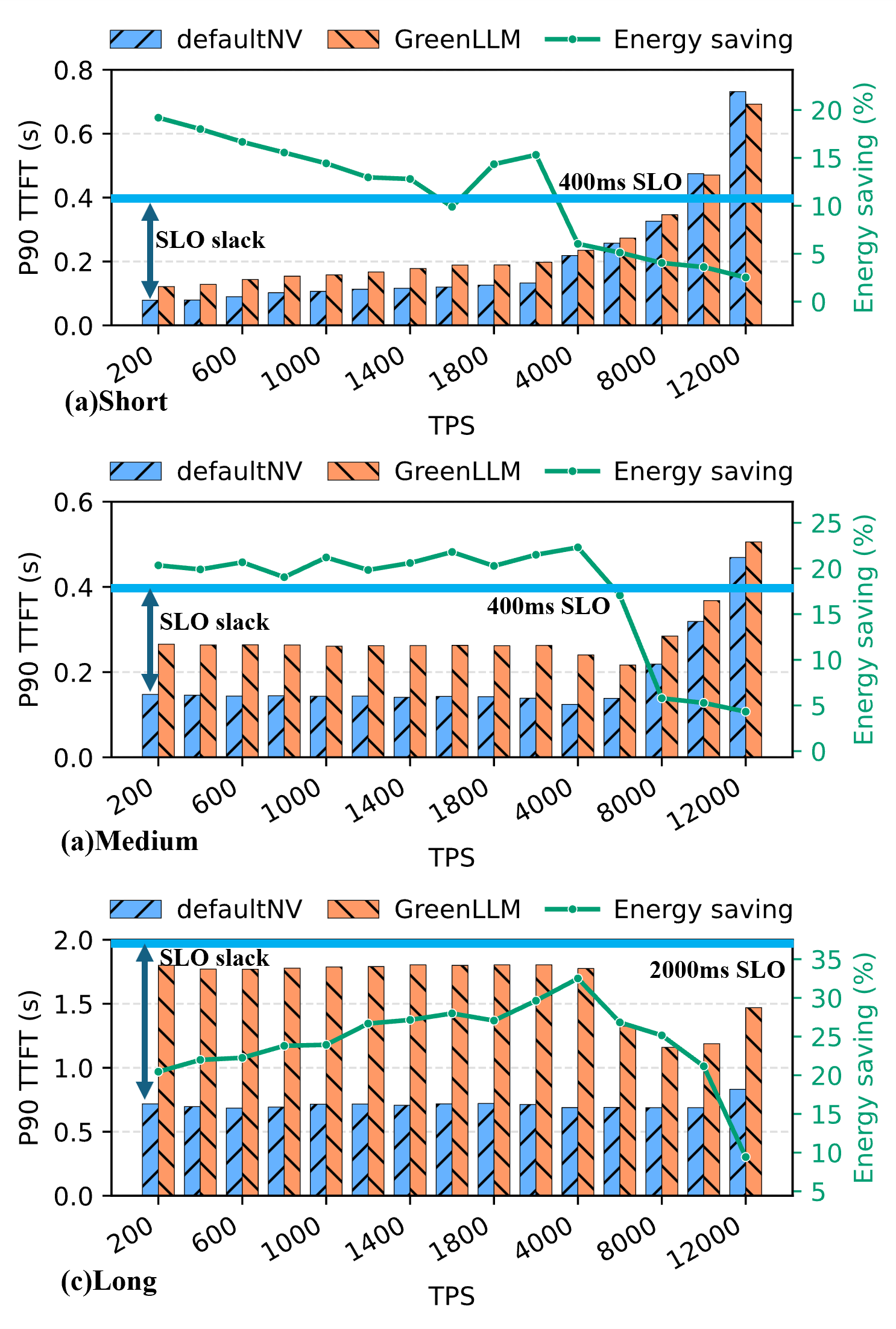}
  \caption{Prefill microbenchmarks (TTFT vs TPS) with defaultNV and GreenLLM. (a) Short; (b) Medium; (c) Long.}
  \label{fig:TTFT_Microbenchmark_Evaluation_prefill}
\end{figure}

\subsubsection{Prefill-Stage Evaluation}
We first evaluate GreenLLM’s prefill optimization under varying prompt lengths and input rates. Figure~\ref{fig:TTFT_Microbenchmark_Evaluation_prefill}a–c present the average time-to-first-token (TTFT) for the Short, Medium, and Long prompt classes across increasing synthetic load (TPS), comparing GreenLLM’s adaptive frequency tuning method against NVIDIA’s default DVFS governor (\textbf{“defaultNV”}). 
We sweep throughput (TPS) to simulate different load levels, measuring P90 TTFT and energy at each point using the microbenchmark as Sect.~\ref{sec:setup}. 
As shown by the line markers in Fig.~\ref{fig:TTFT_Microbenchmark_Evaluation_prefill}a–b, for short/medium prompts (400 ms SLO), GreenDVFS intentionally save energy by exploiting the slack from SLO: TTFT is slightly higher than the default at low–mid TPS, and the energy‑saving curve (green) starts $\sim$10$-$20\% and gradually falls to $\sim$0\%, then collapses near saturation when the controller returns to high clocks to protect the SLO. 
For long prompts (bottom; 2s SLO), the larger prefill cost creates more usable slack: energy savings rise toward $\sim$25–30\% at mid load, and GreenLLM increases the TTFT much higher to save the energy consumption; as TPS climbs further, the slack shrinks. Across all three, TTFT stays within the class SLO through most of the range and energy savings diminish when the system nears saturation.
The microbenchmarks reveal that the defaultNV often achieves TTFT well below the SLO, leaving substantial SLO slack. This slack represents performance headroom that can be traded for energy savings. For instance, on an A100 GPU a moderate-sized request might yield a first-token latency of only $\sim$75ms against a 400 ms SLO, implying 325 ms of slack for possible energy saving.


\begin{figure}[!t]
  \centering
  \includegraphics[width=0.90\linewidth]{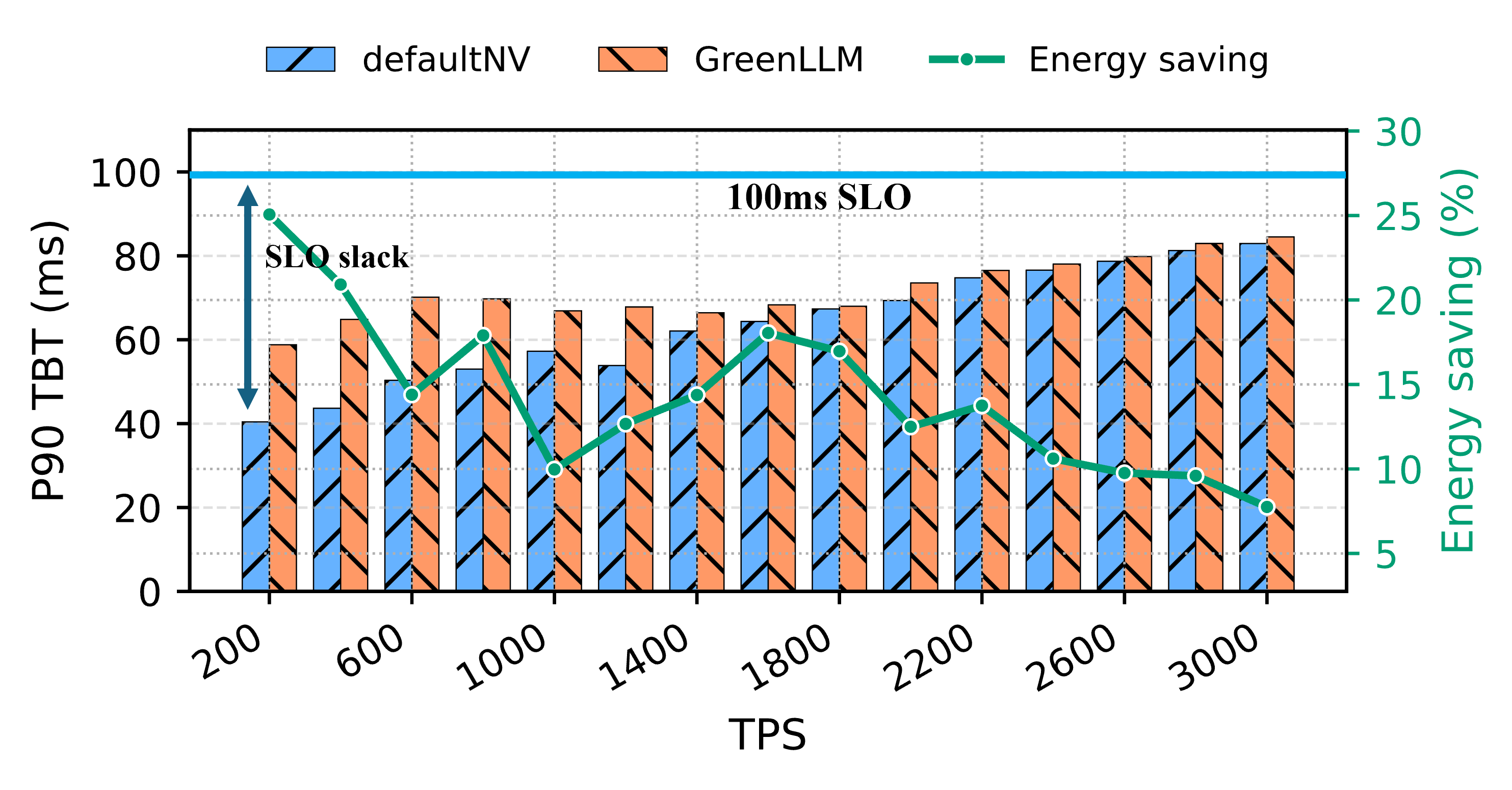}
  \caption{Decode microbenchmarks (TBT vs TPS) with defaultNV and GreenLLM.}
  \label{fig:TBT_Microbenchmark_Evaluation_decode}
\end{figure}

\subsubsection{Decode-Stage Evaluation}
Next, we examine the decoding performance under stable token generation demand, to evaluate GreenLLM’s runtime controller. In this experiment, a fixed number of concurrent decode streams yields a constant aggregate token rate (TPS), which we vary from 200 up to 3000 tokens/s. GreenLLM’s token-tracking DVFS controller is enabled to dynamically adjust the SM frequency every 20 ms, and we compare its results to the \textbf{“defaultNV”}.

Figure~\ref{fig:TBT_Microbenchmark_Evaluation_decode} reports P90 time‑between‑tokens (TBT) across a decode TPS sweep together with GreenLLM’s GPU energy reduction. Across 200–3000 TPS, GreenLLM’s P90 TBT closely tracks the defaultNV and remains within the 100 ms SLO at all points. The gap is most visible only at very light load: at 200–400 TPS GreenLLM shows $\sim$60–66 ms versus $\sim$40–46 ms for the defaultNV, still comfortably below the SLO. As load increases, the difference shrinks—around 1000 TPS we observe $\sim$65 ms (GreenLLM) vs. $\sim$58 ms (defaultNV)—and from 1800 TPS onward the bars are nearly indistinguishable; at 3000 TPS both converge near $\sim$85–86 ms. The energy savings are highest at low TPS ($\sim$20–25\%) and decrease with load, reaching $\sim$8–12\% at 2400–3000 TPS, with a modest rebound around 1600–1800 TPS due to the controller’s band selection. In short, the decode‑phase controller maintains SLO‑compliant token latency while cutting GPU energy by 8–25\%, with the greatest gains when the workload offers more headroom.


\begin{tcolorbox}[takeaway, title=Takeaway \#4]
GreenLLM reduces power while staying close to the SLO target by using a lower frequency. Its scheduler ensures TTFT always meets, but does not exceed, the SLO.
\end{tcolorbox}



\subsubsection{Dynamic-Tracing Evaluation}
\label{subsubsec:dynamic_tracing}
Furthermore, we assess GreenLLM’s ability under a time-varying workload to verify its adaptability and stability. Figure~\ref{fig:TBT_tracing_decode} (in Sect.~\ref{sec:intro}) drives decoding with a synthetic sinusoidal TPS target to test tracking and stability. The top panel shows \textbf{“defaultNV”} (NVIDIA’s scaling governor): the SM clock sits almost stationary in a narrow high band ($\sim$1.1-1.4 GHz) and does not follow changing TPS, confirming the lack of TPS‑aware adaptation. In contrast, the bottom figure shows GreenLLM’s DVFS decisions: the SM clock tracks the workload increase from $\sim$450 MHz to $\sim$1.35 GHz as TPS increases, then symmetrically reverses as TPS falls with small rate-limited adjustments from the 20 ms fine loop and 15 MHz steps. This is consistent with our controller design (Sect.~\ref{subsec:decode-dvfs}): we measure TPS continuously and pick a coarse frequency band, then apply fine, hysteretic nudges to hold TBT within its SLO while minimizing frequency, and thus energy, whenever slack is available. Across the run, p99 TBT stays $\leq$ 100 ms under both policies (GreenLLM 83.2 ms vs. defaultNV 84.6 ms), demonstrating SLO-compliant tracking with lower clocks when demand allows. In the same experiment, GreenLLM achieves 8.9\% lower decode energy while achieving similar or slightly better tail latency (p99 TBT 83.2 ms vs. 84.6 ms).

\begin{tcolorbox}[takeaway, title=Takeaway \#5]
GreenLLM’s dual-loop decode controller design can successfully traces workload intensity and SLO dynamically, modulating SM clocks in real time to save energy.
\end{tcolorbox}

\subsection{Trace Evaluation with Intensity Scaling}
We assess GreenLLM on two production traces, \textbf{Alibaba} and \textbf{Azure}, under multiple intensity levels to capture realistic burstiness and prompt-length skew. We replay the Alibaba trace at \{1, 3, 5, 8, 10\} QPS for at least 30 minutes per run, and We downsample the May 2024 Azure trace to \{1/8, 1/5\} of its original rate to match single-node capacity (the original targets a GPU cluster) while preserving the inter-arrival structure. 
SLOs target \textbf{TTFT} $<$ 400 ms for S/M and $<$ 2 s for L, and \textbf{TBT} $<$ 100 ms during decoding~\cite{stojkovic2025dynamollm}. We collect the pass rates (TTFT\%, TBT\%), together with energy for prefill and decode. We compare three configurations: DefaultNV, PrefillSplit, and GreenLLM introduced in Sect.~\ref{subsec:models_traces}. All experiments are run with two representative LLMs: a dense model Qwen3‑14B and a mixture‑of‑experts model, Qwen3‑30B‑MoE, introduced in Table~\ref{tab:models}.

Table~\ref{tab:q14-chat-energy-rel} shows results with Qwen3-14B. Energy savings of GreenLLM drop from 27.5\% at 1 QPS to 6.8\% at 10 QPS as higher decode clocks and power are needed to sustain TBT. On Azure, however, GreenLLM consistently saves 28–34\%, mainly from reduced decode energy (0.62–0.73$\times$ defaultNV). Similar trends hold for Qwen3-30B-MoE in Table~\ref{tab:q30-chat-energy-rel}: total energy falls by 19.7–31\% on Azure and 10–21\% on Alibaba, with decode energy at 0.73–0.89$\times$ defaultNV.


\begingroup
\setlength{\tabcolsep}{3pt}%
\sisetup{table-number-alignment=center,table-text-alignment=center}

\begin{table}[t]
\caption{Energy and SLOs on chat workloads from Qwen-14B (energies normalized to defaultNV}
\label{tab:q14-chat-energy-rel}
\centering
\scriptsize
\begin{tabular*}{\columnwidth}{@{\extracolsep{\fill}} l l
                S[table-format=1.3]
                S[table-format=1.3]
                S[table-format=3.1]
                S[table-format=3.1]
                S[table-format=3.2]@{}}
\toprule
Workload & Method & {Rel.\ Decode} & {Rel.\ Prefill} & {TTFT (\%)} & {TBT (\%)} & {$\Delta$En (\%)} \\
\midrule
  & defaultNV      & 1.000 & 0.589 & 99.3 & 98.2 & 0.00 \\
            chat\_1qps & PrefillSplit & 0.998 & 0.586 & 98.5 & 100.0 & 0.28 \\
            & GreenLLM     & 0.675 & 0.479 & 98.9 & 99.6  & 27.52 \\
\addlinespace[2pt]
  & defaultNV      & 1.000 & 0.903 & 99.5 & 98.6 & 0.00 \\
            chat\_3qps & PrefillSplit & 0.995 & 0.878 & 98.8 & 98.6 & 1.61 \\
            & GreenLLM     & 0.798 & 0.714 & 99.2 & 95.1 & 20.53 \\
\addlinespace[2pt]
  & defaultNV      & 1.000 & 1.072 & 96.0 & 98.6 & 0.00 \\
            chat\_5qps & PrefillSplit & 1.006 & 1.046 & 99.0 & 98.6 & 0.94 \\
            & GreenLLM     & 0.832 & 0.865 & 98.2 & 96.0 & 18.13 \\
\addlinespace[2pt]
  & defaultNV      & 1.000 & 1.314 & 89.9 & 96.1 & 0.00 \\
            chat\_8qps & PrefillSplit & 1.023 & 1.292 & 96.4 & 96.4 & -0.04 \\
            & GreenLLM     & 0.974 & 1.083 & 95.7 & 95.1 & 11.16 \\
\addlinespace[2pt]
 & defaultNV      & 1.000 & 1.213 & 84.9 & 90.7 & 0.00 \\
            chat\_10qps & PrefillSplit & 1.008 & 1.183 & 89.1 & 91.3 & 1.03 \\
            & GreenLLM     & 1.029 & 1.034 & 88.2 & 90.9 & 6.78 \\
\addlinespace[2pt]
  & defaultNV      & 1.000 & 1.707 & 98.5 & 100.0 & 0.00 \\
              Azure\_code5& PrefillSplit & 0.964 & 1.663 & 96.6 & 100.0 & 2.95 \\
              & GreenLLM     & 0.629 & 1.294 & 94.2 & 100.0 & 28.98 \\
\addlinespace[2pt]
  & defaultNV      & 1.000 & 1.696 & 99.9 & 100.0 & 0.00 \\
              Azure\_code8& PrefillSplit & 0.981 & 1.652 & 100.0 & 100.0 & 2.35 \\
              & GreenLLM     & 0.728 & 1.129 & 99.9  & 100.0   & 31.12 \\
\addlinespace[2pt]
  & defaultNV      & 1.000 & 1.393 & 99.7 & 100.0 & 0.00 \\
              Azure\_conv5 & PrefillSplit & 0.992 & 1.370 & 100.0 & 100.0 & 1.32 \\
              & GreenLLM     & 0.624 & 0.954 & 99.8 & 100.0 & 34.09 \\
\addlinespace[2pt]
  & defaultNV      & 1.000 & 1.416 & 100.0 & 100.0 & 0.00 \\
             Azure\_conv8 & PrefillSplit & 0.993 & 1.373 & 100.0 & 100.0 & 2.07 \\
              & GreenLLM     & 0.713 & 0.922 & 100.0 & 100.0 & 32.31 \\
\bottomrule
\end{tabular*}
\end{table}
\endgroup

\begingroup
\setlength{\tabcolsep}{3pt}%
\sisetup{table-number-alignment=center,table-text-alignment=center}
\begin{table}[t]
\caption{Energy and SLOs on chat workloads from Qwen-30B (MoE) (energies normalized to defaultNV}
\label{tab:q30-chat-energy-rel}
\centering
\scriptsize
\begin{tabular*}{\columnwidth}{@{\extracolsep{\fill}} l l
                S[table-format=1.3]
                S[table-format=1.3]
                S[table-format=3.1]
                S[table-format=3.1]
                S[table-format=3.2]@{}}
\toprule
Workload & Method & {Rel.\ Decode} & {Rel.\ Prefill} & {TTFT (\%)} & {TBT (\%)} & {$\Delta$En (\%)} \\
\midrule
\addlinespace[2pt]
           & defaultNV      & 1.000 & 0.650 & 100.0 & 100.0 & 0.00 \\
                   chat\_1qps & PrefillSplit & 1.010 & 0.643 & 100.0 & 100.0 & -0.16 \\
                    & GreenLLM     & 0.731 & 0.580 & 100.0 & 99.6 & 20.56 \\
\addlinespace[2pt]
           & defaultNV      & 1.000 & 0.790 & 99.4 & 97.3 & 0.00 \\
                   chat\_3qps & PrefillSplit & 1.001 & 0.784 & 99.2 & 97.3 & 0.30 \\
                    & GreenLLM     & 0.855 & 0.712 & 99.2 & 95.6 & 12.46 \\
\addlinespace[2pt]
           & defaultNV      & 1.000 & 0.872 & 99.3 & 97.9 & 0.00 \\
                  chat\_5qps  & PrefillSplit & 1.004 & 0.834 & 99.1 & 98.0 & 1.82 \\
                    & GreenLLM     & 0.890 & 0.791 & 99.3 & 96.2 & 10.23 \\
                    
 & defaultNV      & 1.000 & 1.063 & 99.9 & 99.6 & 0.00 \\
                   Azure\_conv5 & PrefillSplit & 0.995 & 1.051 & 99.7 & 99.7 & 0.82 \\
                    & GreenLLM     & 0.808 & 0.849 & 99.7 & 96.1 & 19.67 \\
\addlinespace[2pt]
 & defaultNV      & 1.000 & 0.921 & 99.9 & 99.8 & 0.00 \\
                  Azure\_conv8  & PrefillSplit & 0.997 & 0.908 & 100.0 & 99.7 & 0.88 \\
                    & GreenLLM     & 0.783 & 0.706 & 100.0 & 93.8 & 22.51 \\
\addlinespace[2pt]
 & defaultNV      & 1.000 & 1.140 & 100.0 & 99.7 & 0.00 \\
                   Azure\_code5 & PrefillSplit & 0.993 & 1.112 & 99.9 & 99.8 & 1.57 \\
                    & GreenLLM     & 0.642 & 0.883 & 100.0 & 95.9 & 28.76 \\
\addlinespace[2pt]
 & defaultNV      & 1.000 & 0.985 & 99.6 & 99.8 & 0.00 \\
                  Azure\_code8  & PrefillSplit & 1.001 & 0.952 & 100.0 & 100.0 & 1.61 \\
                    & GreenLLM     & 0.655 & 0.713 & 99.6 & 97.0 & 31.05 \\

\bottomrule
\end{tabular*}
\end{table}
\endgroup

Across traces and intensities, \textbf{GreenLLM} maintains high SLO pass rates. For Alibaba chat, Table~\ref{tab:q14-chat-energy-rel} shows that TTFT\%, TBT\% remain $\ge$95\% through 1$-$8 QPS.
Especially for QPS 8, the TTFT\% from \textbf{defaultNV} shows much worse result, where it is only 89.9\% compared to 95.7\% in GreenLLM.
At 10 QPS, GreenLLM's pass rates drop to TTFT 88.2\% and TBT 90.9\%—consistent with the system nearing prefill saturation, at which point the decode optimizer raises clocks to protect streaming quality.

On Azure code (Qwen3‑14B), rest part of Table~\ref{tab:q14-chat-energy-rel}, \textbf{GreenLLM} retains TBT 100\% while TTFT\% varies with long‑prompt mix (e.g., 94.2\% on \texttt{code5}); conversation workloads maintain $\sim$99–100\% for both TTFT and TBT. For Qwen3‑30B‑MoE, demonstrated in Table~\ref{tab:q30-chat-energy-rel}, pass rates are similarly high at feasible intensities with occasional modest TBT\% reductions (e.g., 93.8–97.0\%) on Azure code slices.

The routing‑only ablation \textbf{PrefillSplit} reduces head‑of‑line interference and slightly tightens TTFT tails, but yields only $\lesssim$1–3\% energy change across traces and models. In contrast, \textbf{GreenLLM}’s \emph{phase‑aware DVFS} (prefill optimizer + decode optimizer) delivers the decisive savings: on Azure with Qwen3‑14B, decode energy falls to 0.63–0.71$\times$ Default; with Qwen3‑30B‑MoE it falls to 0.64–0.80$\times$, all while sustaining high TTFT/TBT pass rates (Tables~\ref{tab:q14-chat-energy-rel}$-$\ref{tab:q30-chat-energy-rel}). These results validate that exploiting the prefill/decode split, and controlling frequency using dedicated strategy for prefill/decode, is the right lever for energy proportionality under SLOs.

\begin{tcolorbox}[takeaway, title=Takeaway \#6]
Across all model families and traces, GreenLLM consistently reduces GPU energy at scale without sacrificing throughput and with high SLO compliance.
\end{tcolorbox}

\subsection{SLO Margin Sensitivity: Energy–SLO Tradeoffs}
We next evaluate how the latency SLO margin, the slack or tightness around target latencies, affects GreenLLM’s energy–latency tradeoffs. In this experiment, we keep these SLO targets by scaling them with margin factors (as described at Sect.~\ref{sec:greenLLM}, for prefill, the margin factor is coefficient applied on latency deadline D) ranging from an extremely strict 0.2× up to a highly relaxed 2.0× (with intermediate values 0.6×, 0.85×, 0.95×, and 1.2×), and measure the impact on GPU energy consumption and tail latency (90th percentile). Here we use the Alibaba chat trace (10 QPS) on Qwen-14B model.

\begin{figure}[!t]
  \centering
  \begin{subfigure}{\columnwidth}
    \centering
    \includegraphics[width=0.85\linewidth]{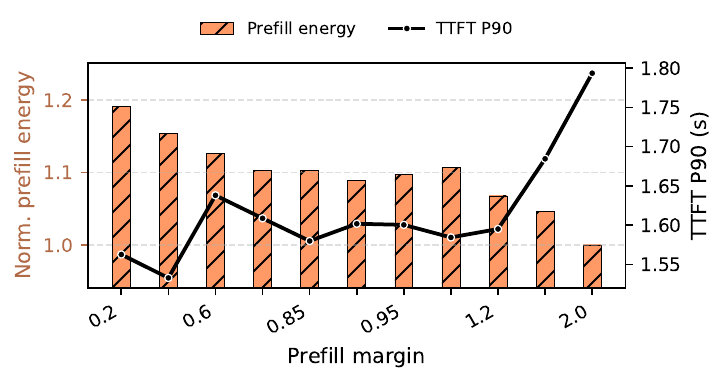}
    \caption{Prefill margin with fixed Decode margin $x0.95$.}
    \label{fig:prefill-sub}
  \end{subfigure}


  \begin{subfigure}{\columnwidth}
    \centering
    \includegraphics[width=0.85\linewidth]{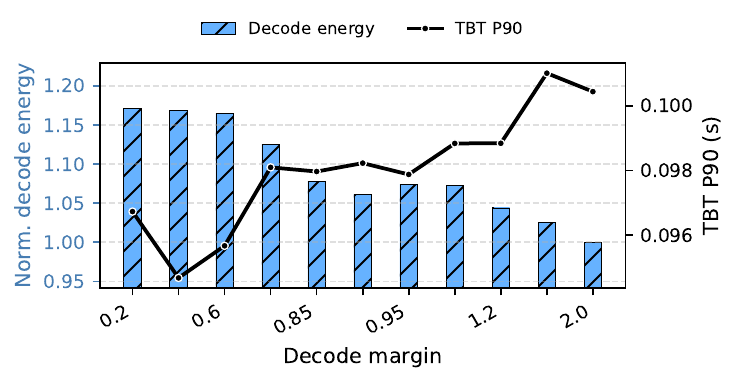}
    \caption{Decode margin with fixed prefill margin $x0.95$.}
    \label{fig:decode-sub}
  \end{subfigure}

  \caption{Margin sensitivity: energy–latency tradeoffs in prefill and decode.}
  \label{fig:margin-sensitivity}
\end{figure}

We first adjust the TTFT margin factor while holding the decode-phase margin at a representative margin of 0.95× (nearly the baseline target). Figure~\ref{fig:prefill-sub} illustrates the total energy consumed in the prefill stage (lines) and the achieved P90 TTFT (bars) for each margin setting. As expected, looser margin reduce energy usage and increase the TTFT. At the strict end (e.g., margin 0.6×, a 40\% tighter deadline), GreenLLM’s prefill controller must raise clock frequencies to speed up computation processing, showing a much lower P90 TTFT, however, result in higher prefill energy consumption. 
Conversely, looser prefill margin yield substantial energy savings. At a relaxed margin like 1.2× (20\% more slack), or even at 2.0×, which doubles the latency allowance, the prefill controller can dial down GPU utilization while almost violate latency requirement. Under these high-margin settings, the P90 TTFT increases, for example, rising from less than $\sim$1000 ms at baseline to almost $\sim$1640 ms at 1.2×, staying well within the allowed SLO buffer. 

Furthermore, we perform a similar experiment for the decode stage, varying the TBT margin while fixing the prefill margin at 0.95×. Figure~\ref{fig:decode-sub} plots the total decode energy (bars) and P90 TBT (lines) as we sweep the decode margin from tight (0.2×) to relaxed (2.0×). The trends here closely mirror the prefill-phase behavior. Tightening the decode budget forces GreenLLM to increase GPU frequency during token generation, which drives up energy usage. 
For example, at a strict 0.85× decode margin, the controller ramps up to maintain a faster TPS, and P90 TBT improves to meet the $\sim$85–90 ms target with no SLO violations. In contrast, a relaxed decode SLO (e.g., 1.2× or 2.0×) saves more energy at lower clocks. The 90th-percentile token latency grows only slightly under these looser settings (reaching $\sim$110–120 ms at 1.2×), and violates the acceptable SLO for streaming outputs.
These margin-sensitivity experiments demonstrate that GreenLLM automatically tunes its behavior to both stricter and more lenient latency requirements, enabling a smooth energy–latency tradeoff without any manual reconfiguration. 

\begin{tcolorbox}[takeaway, title=Takeaway \#7]
GreenLLM preserves SLO compliance across diverse latency budgets, achieving SLO-aware energy optimization automatically without re-engineering.
\end{tcolorbox}


%% file: s7_stoa.tex
\section{Related Work}

\subsection{LLM Disaggregation and Scheduling}
LLM inference consists of a compute-intensive prefill phase and a sequential, memory-bound decode phase, whose differing resource profiles complicate efficient serving~\cite{agrawal2024taming,zhong2024distserve,vllm_docs_optimization,nvidia_inference_optimization_2023,kwon2023pagedattention}. Splitwise~\cite{patel2024splitwise} addresses this by disaggregating the two phases across distinct “prompt” and “token” machines, transferring KV caches over high-speed links to exploit heterogeneous clusters and improve throughput and cost efficiency. Other scheduling techniques instead focus on the decode phase: Orca~\cite{yu2022orca} introduced in-flight batching to insert new requests into ongoing decode iterations, improving utilization and latency. More recently, Jaillet et al. modeled decode scheduling under GPU memory constraints and proposed online algorithms that reduce latency and energy by better managing prompt lengths and KV cache usage~\cite{jaillet2025online}. These efforts primarily optimize batching and machine allocation for throughput or latency. In contrast, GreenLLM manages phases within a single GPU, combining prompt-length–aware queueing and phase-specific DVFS rather than relying on static phase placement or batching alone.

\subsection{GPU Power Management}
DVFS is a central mechanism for trading performance and energy, and has been studied extensively for ML workloads. Nabavinejad et al.~\cite{nabavinejad2022coordinated} coordinated batching delays with GPU DVFS to save energy in CNN inference, although such coarse-grained control is poorly suited to unpredictable LLM decode loops. Wang et al.~\cite{wang2025asplos_dvfs} showed that Huawei’s Ascend NPUs support millisecond-level frequency control, enabling per-operator DVFS guided by accurate power–performance models; this fine-grained tuning reduced core power by over 13\% with negligible slowdown. Patel et al.~\cite{patel2024characterizing} characterized the usage of LLM inference power and built POLCA, which oversubscribes GPU power budgets at cluster scale to host more LLM servers under datacenter limits. Wilkins et al.~\cite{wilkins2024offline} developed workload-level energy models parameterized by prompt and output lengths, and used them to design an offline energy-optimal heterogeneous scheduler. Together, these works demonstrate the value of analytical models and adaptive DVFS at both the operator and cluster levels. GreenLLM adopts a complementary approach: builds compact latency-power models online for the prefill phase, and applies queueing-aware optimization to set GPU frequencies dynamically, extending model-driven DVFS to runtime per-phase control on commodity GPUs.

\subsection{SLO-Aware Energy Optimization}
The scale of LLM deployment has made inference the dominant contributor to lifecycle energy, motivating techniques that reduce consumption while meeting latency SLOs. Kakolyris et al.~\cite{kakolyris2024slo} proposed an iteration-level DVFS controller that tracks the decode loop and adjusts the GPU frequency to maintain 99th percentile latency targets, producing up to 45\% energy savings. An extended version, throttLL’eM~\cite{kakolyris2025throttll}, further incorporated KV-cache projections and autoscaling to cut cluster energy by over 40\% under latency bounds. Complementary work has quantified the environmental footprint of inference, showing that large models can consume tens of watt-hours per query, emphasizing the urgency of runtime optimizations~\cite{kakolyris2024slo,kakolyris2025throttll}. Beyond DVFS, model- or cluster-level schedulers such as ClockWork~\cite{gujarati2020serving}, and DynamoLLM~\cite{stojkovic2025dynamollm} coordinate batching, autoscaling, and frequency scaling to improve cost and performance efficiency, though they do not exploit LLMs’ two-phase structure. Compared to these efforts, GreenLLM integrates SLO awareness directly into phase-specific control: prompt-length–based queuing reduces time-to-first-token, latency–power modeling informs prefill DVFS, and a token-tracking hysteresis controller adapts decode frequency. This unified design enables substantial energy savings while respecting stringent service-level objectives.